\documentclass[11pt]{article}
\usepackage{slashed}
\usepackage{amsmath}
\usepackage{amssymb}
\usepackage{bbm}
\usepackage{epsfig}
\allowdisplaybreaks[4]
\begin{document}
\title{Can we build a sensible theory with broken charge and colour symmetries?}
\author{P.M. Ferreira \\ CFTC, Faculdade de Ci\^encias,\\ 
Universidade de Lisboa, Portugal}
\date{March, 2002} 
\maketitle
\noindent
{\bf Abstract:} Charge and colour symmetries are broken using the vevs of scalar
fields, but it is shown the theory possesses a photon-like massless particle and
electroweak interactions almost identical to those of the Standard Model. We
consider experimental signatures of this theory. 
\vspace{-9cm}
\vspace{10cm}

\section{Introduction}
In the Standard Model (SM) gauge symmetry breaking is achieved through a 
non-zero vacuum expectation value (vev) acquired by a scalar field. This field
has no colour or electric charge. So, the QCD and electromagnetic symmetries,
$SU(3)\times U(1)$, are preserved and we are left with massless gluons and 
photons. However, in supersymmetric models there are several other scalars,
partners of the fermionic fields, which have colour and electric charge.
It is then possible that there are minima of the scalar potential with non-zero 
vevs for some charged or coloured fields. The requirement that the lowest of 
these minima does not have this type of vevs was first explored by Fr\'ere {\em 
et al}~\cite{fre} to constrain the extensive parameter space of supersymmetric
models. The study of these charge and colour breaking (CCB) minima was further 
developed by many authors~\cite{eve}.

In this article we wish to explore an alternative point of view: is it possible 
to construct models where charged and coloured fields have non-zero vevs and, at
the same time, obtain a massless gauge field that resembles the normal photon? 
In fact, we build such a model and show that its electroweak sector is almost 
identical to the SM's. Nevertheless, the ``new" photon and Z boson have also 
some mixing with the $SU(3)$ gauge fields. In the strong sector, the differences
with the SM are larger - five gluons are massive, four of these have electric 
charge and the quarks have integer electric charge, which depends on their 
colour. Hence we obtain, in this context, the quark charge assignments of the 
Han-Nambu model~\cite{han}. Models with Han-Nambu quarks have been obtained 
before from the spontaneous symmetry breaking of gauge groups larger than the 
SM's~\cite{pati}. In our approach, we start we the SM gauge group and the 
integer charged quarks arise because the colour group $SU(3)$ is broken.

This article is structured as follows: in the next section we build the 
model and analyse the gauge bosons' masses, mixings, their interactions with
fermions and self-couplings. In section~\ref{sec:test} we compare the 
predictions from this model with the SM ones, with special emphasis put on 
results from two-photon physics. These will be shown to be better explained by a
theory with integer charged quarks. We close with a general discussion and
conclusions. 

\section{Model building} 
\label{sec:mod}
Our starting point is the gauge group $SU(3) \times SU(2) \times U(1)$ which 
will be spontaneously broken to $SU(2) \times U(1)$ via the Higgs mechanism. 

\subsection{Breaking the gauge symmetry}

Let us consider a general theory with a gauge group ${\cal G}$ with 
$N$ generators $T^a$ and $n$ scalar fields $\phi_i$, each with a vacuum 
expectation value described by the vector $|v_i>$. Then, as is well known, 
the covariant derivatives are given by
\begin{equation}
D_\mu \phi_i \; = \; \partial_\mu \phi_i \,+\, \frac{i}{2}\,\sum_{a=1}^N\,g_a\,
T^a \,A^a_\mu\,\phi_i \;\;\; ,
\end{equation} 
where $g_a$ is the gauge coupling corresponding to the generator $T^a$ and 
$A^a_\mu$ its respective gauge field. When the fields $\phi_i$ acquire vevs,
$|v_i>$, mass terms for the gauge fields are produced with a mass matrix given 
by (no sum in $\{a,b\}$)
\begin{equation}
M^2_{ab} \; =\; \frac{g_a\,g_b}{2}\,\sum_{i=1}^n \, <v_i|\,T^a\,T^b + T^b\,T^a\,
|v_i> \;\;\; .
\label{eq:mass}
\end{equation}
In the SM there is a single $SU(2)$ doublet scalar field. This produces 
the known spectrum of gauge boson masses: massive W's and $Z^0$, massless gluons
and photons. 

Let us now consider the Minimal Supersymmetric Standard Model 
(MSSM). Gauge symmetry breaking is achieved by two colourless Higgs doublets 
$H_1$ and $H_2$ with vevs $v_1/\sqrt{2}$ and $v_2/\sqrt{2}$ respectively. Among 
the many possible ways of causing CCB (see ref.~\cite{cas}
for a detailed review) we have chosen the case where the scalar partners of the 
top quark $\{t_L \, , \, t_R\}$ acquire non-zero vevs. The field $t_L$ is an 
$SU(3)$ triplet which is part of the $SU(2)_L$ doublet $Q_L = (t_L \;,\; b_L)$, with hypercharge $y_Q = 1/6$. The field $t_R$ is a $\bar{3}$ of $SU(3)$ and 
singlet of $SU(2)_L$ with hypercharge $y_t = -2/3$. We denote by $q/\sqrt{2}$ 
and $t/\sqrt{2}$ the vevs of $t_L$ and $t_R$ respectively, {\em both with colour
index 3}. This means we are effectively breaking the colour symmetry along a 
particular direction. In the Standard Model these vevs would have an  
electric charge $\pm 2/3$ and thus, in principle, the electromagnetic gauge 
invariance would be broken. However we will see that this is not the case and 
the model still maintains an unbroken $U(1)$ subgroup. 

Because we are considering vevs carrying a particular colour index, 
it is convenient to alter slightly the usual notation: let $Q_L$ be written as 
\begin{equation}
Q_L \; = \; \begin{pmatrix} t_L \\ b_L \end{pmatrix} \; \equiv \; 
\begin{pmatrix} t_L^1 \vspace{0.2cm} \\ t_L^2 \vspace{0.2cm} \\ t_L^3 
\vspace{0.2cm} \\ b_L^1 \vspace{0.2cm} \\ b_L^2 \vspace{0.2cm} \\ b_L^3 
\vspace{0.2cm} \end{pmatrix}
\; \begin{array}{c} \mbox{vacuum} \\ \longrightarrow \\ \end{array} 
\begin{pmatrix} 0 \\ 0 \\ q/\sqrt{2} \\ 0 \\ 0 \\ 0  \end{pmatrix} \;\;\; .
\end{equation}
We write the $Q_L$ covariant derivative as
\begin{equation}
D_\mu Q_L \; = \; \partial_\mu Q_L \;+\; i\,g^{\prime}\,Y\,B_\mu \;+\; 
\frac{i}{2}\,g\,\sigma^{\prime}_i\,Q_L\,W_\mu^i \;+\; \frac{i}{2}\,g_3\,
\lambda^{\prime}_a\,Q_L\,G^a_\mu \;\;\; ,
\label{eq:der}
\end{equation}
where $\{g^{\prime} , g , g_3\}$ are the $U(1)$, $SU(2)$ and $SU(3)$ gauge 
couplings  and $\{B_\mu, W_\mu^i , G^a_\mu\}$ ($i = 1, 2, 3$, $a = 1, \ldots 8$)
are their respective gauge fields. In this representation the hypercharge matrix
$Y$ is, for a field of hypercharge $y$, given by $Y = y\,
\mathbbm{1}_6$, and the generalised Pauli and Gell-Mann matrices 
$\sigma^{\prime}_i$, $\lambda^{\prime}_a$ are
\begin{align}
\sigma^{\prime}_1 \; = & \begin{pmatrix} 0 & \mathbbm{1}_3 \\ \mathbbm{1}_3 & 0
\end{pmatrix} & \sigma^{\prime}_2 \; = & \begin{pmatrix} 0 & -i\,\mathbbm{1}_3 
\\ i\,\mathbbm{1}_3 & 0 \end{pmatrix} & 
\sigma^{\prime}_3 \; = & \begin{pmatrix} \mathbbm{1}_3 & 0 \\ 0 & \mathbbm{1}_3 
\end{pmatrix} & \lambda^{\prime}_a  \; = & \begin{pmatrix} \lambda_a & 0 \\ 0 & 
\lambda_a \end{pmatrix} \;\;\; ,
\end{align}
with $\mathbbm{1}_n$ the $n\times n$ unit matrix and $\lambda_a$ the usual
Gell-Mann matrices. Likewise, for the field $t_R$, we have
\begin{equation}
t_R \; = \; \begin{pmatrix} t_R^1 \vspace{0.2cm} \\ t_R^2 \vspace{0.2cm} \\ 
t_R^3 \vspace{0.2cm} \end{pmatrix} \; 
\begin{array}{c} \mbox{vacuum} \\ \longrightarrow \\ \end{array} 
\begin{pmatrix} 0 \\ 0 \\ t/\sqrt{2} \end{pmatrix} \;\;\; ,
\end{equation}
and its covariant derivative has an expression similar to eq.~\eqref{eq:der}, 
without the $g$ term and with the matrices $\lambda_a$ replaced by 
$-\lambda_a^*$. At this point we can use eq.~\eqref{eq:mass} and
write down the gauge bosons' mass matrices. Since the Gell-Mann matrices $\{
\lambda_1, \lambda_2, \lambda_3\}$ have zeros in their third row and column 
the gluons $G^1_\mu$, $G^2_\mu$ and $G^3_\mu$ remain massless. On the other hand
$G^4_\mu$ to $G^7_\mu$ acquire a mass given by
\begin{equation}
M^2_{G_{4\ldots 7}} \; = \; \frac{1}{4}\,g_3^2\,(q^2\,+\,t^2) \;\;\; .
\end{equation}
Our choice of colour breaking implies that there is no mixing among any of 
these seven fields, or between them and the electroweak gauge fields. Like in 
the SM the fields $W^1_\mu$ and $W^2_\mu$ mix to form the charged weak bosons 
$W^\pm$ with mass given by  
\begin{equation}
M^2_W \; = \; \frac{1}{4}\,g^2\,(v^2\,+\,q^2)\;\;\; ,
\end{equation}
with $v^2 = v_1^2\,+\,v_2^2$. Finally the only non-trivial mass matrix is the
one involving the gauge bosons $W^3_\mu$, $B_\mu$ and $G^8_\mu$, 
namely~\footnote{This study is valid for the case of $m$ scalar 
higgs-like doublets each with a different vev $v_i$, simply replacing $v^2$ by $
{\displaystyle \sum_{i=1}^m \, v_i^2}$ in the formulae.}, 
\begin{equation}
[M^2_{G^0}] \; = \; \begin{pmatrix} {\displaystyle \frac{g^2}{4}\,(v^2\,+\,q^2)}
& -\,{\displaystyle \frac{g\,g^\prime}{4}\,(v^2\,-\,2\,y_Q\,q^2)} & -\,
{\displaystyle \frac{g\,g_3}{2\,\sqrt{3}}\, q^2} \\
-\,{\displaystyle \frac{g\,g^\prime}{4}\,(v^2\,-\,2\,y_Q\,q^2)} & 
{\displaystyle\frac{{g^\prime}^2}{4}\,(v^2\,+\,4\,y_Q^2\,q^2\,+\,4\,y_t^2\,t^2)}
& -\,{\displaystyle \frac{g^\prime\,g_3}{\sqrt{3}}\,(y_Q\,q^2\,-\,y_t\,t^2)} \\ 
-\,{\displaystyle \frac{g\,g_3}{2\,\sqrt{3}}\,q^2} & -\,{\displaystyle \frac{
g^\prime\,g_3}{\sqrt{3}}\,(y_Q\,q^2\,-\,y_t\,t^2)} & {\displaystyle \frac{
g_3^2}{3}\,(q^2\,+\,t^2)} \end{pmatrix} \;\;\; .
\label{eq:mg0}
\end{equation}
The determinant of this matrix is 
\begin{equation}
\det[M^2_{G^0}] \; = \; \frac{1}{12}\,\left(y_Q\,+\,y_t\,+\,\frac{1}{2}\right)^2
\, {g^\prime}^2\,g^2\,g_3^2\,v^2\,q^2\,t^2 \;\;\; .
\end{equation}
If either $q$ or $t$ are zero we immediately have a massless neutral gauge 
boson. Obviously, $q = t = 0$ is the SM case. However, it is interesting to 
point out that the quantum numbers of $\{t_L \,,\,t_R\}$ in the MSSM are
{\em exactly} such that this determinant is zero - with the values of the 
hypercharges given before we have $y_Q \,+\,y_t\,=\,-1/2$. The relevance of 
obtaining a massless particle will be evident in the next chapter, but we can 
anticipate that, despite the fact it results from mixing with a gluon, this 
particle has interactions that make it virtually identical to the photon. 

The eigenvalues of matrix~\eqref{eq:mg0} are $0$ and $(A \pm \sqrt{ 
A^2-4\,B})/2$, with 
\begin{eqnarray}
A & = & \frac{{g^\prime}^2\,+\,g^2}{4}\,v^2\,+\, \left( \frac{g^2}{4}\,+\,
{g^\prime}^2\,y_Q^2\,+\,\frac{g_3^2}{3}\right)\,q^2\,+\,\left[\left(y_Q\,+\,
\frac{1}{2}\right)^2\,{g^\prime}^2\,+\,\frac{g_3^2}{3}\right]\,t^2 \nonumber \\
B & = & \frac{1}{4}\,\left[\frac{g_3^2}{3}\,({g^\prime}^2\,+\,g^2)\,+\,
\frac{1}{4}\,\left(y_Q\,+\,\frac{1}{2}\right)^2\,{g^\prime}^2\,g^2\right]\,
\left[q^2\,t^2\,+\,v^2\,(q^2\,+\,t^2)\right] \;\;\; .
\label{eq:m2}
\end{eqnarray}
We want to identify one of these massive eigenvalues with the $Z$ boson, denoted
by $\tilde{Z}$. The other one which, for physical reasons to be explained 
shortly, will be heavier than the $Z$, we call $\tilde{G}$. The eigenstates 
$\tilde{A}_\mu$, corresponding to the zero eigenvalue, $\tilde{Z}_\mu$ and 
$\tilde{G}_\mu$ are given by the unitary transformation
\begin{align}
\tilde{A}_\mu &= a_1\,B_\mu \,+\,b_1\,W^3_\mu \,+\,c_1\,G^8_\mu \nonumber \\
\tilde{Z}_\mu &= a_2\,B_\mu \,+\,b_2\,W^3_\mu \,+\,c_2\,G^8_\mu \nonumber \\
\tilde{G}_\mu &= a_3\,B_\mu \,+\,b_3\,W^3_\mu \,+\,c_3\,G^8_\mu \;\;\; ,
\label{eq:mix}
\end{align}
with
\begin{eqnarray}
\frac{b_1}{a_1} & = & \frac{g^\prime}{g} \nonumber \\
\frac{c_1}{a_1} & = & \frac{\sqrt{3}}{2}\,\frac{g^\prime}{g_3}\,(2\,y_Q\,+\,1)
\nonumber \\
\frac{b_{2,3}}{a_{2,3}} & = & -\,\frac{g}{g^\prime}\,\frac{(2\,y_Q\,+\,1)\,[2\,
y_Q\,q^2 \,+\,(2\,y_Q\,+\,1)\,t^2]\,{g^\prime}^2\,{\displaystyle \frac{4}{3}}\,
g_3^2\,(q^2\,+\,t^2)\;-\;4\,m_{2,3}^2}{(2\,y_Q\,+\,1)\,g^2\,q^2\,+\,
{\displaystyle \frac{4}{3}}\,g_3^2\,(q^2\,+\,t^2)\;-\;4\,m_{2,3}^2} \nonumber \\
\frac{c_{2,3}}{a_{2,3}} & = & \frac{2\,\sqrt{3}}{3}\,\frac{g_3}{g^\prime}\,
\frac{(2\,y_Q\,{g^\prime}^2\,-g^2)\,q^2\,+\,(2\,y_Q\,+\,1)\,{g^\prime}^2\,t^2}{
(2\,y_Q\,+\,1)\,g^2\,q^2\,+\,{\displaystyle \frac{4}{3}}\,g_3^2\,(q^2\,+\,t^2)\;
-\;4\,m_{2,3}^2} \;\;\;.
\label{eq:coef}
\end{eqnarray}
Notice that the coefficients corresponding to the 
massless field depend only on the gauge couplings, as opposed to the remaining 
two, which depend on the vevs. Clearly the SM limit cannot be obtained from 
eq.~\eqref{eq:coef} by taking the limit $q \rightarrow 0$, $t \rightarrow 0$. In
fact in this limit the initial matrix~\eqref{eq:mg0} has a two-fold degenerate
zero eigenvalue. 

\subsection{Gauge interactions of fermions}
\label{sec:int}

We obtain the gauge interactions of a fermionic field $F$ by replacing, in the
free lagrangean, partial derivatives by covariant ones, so that
\begin{equation}
\bar{F}\,\slashed{\partial}\,F\; \rightarrow\; \bar{F}\,\slashed{\partial}\,F\,
+\,i\,g^\prime\,y_F\,\bar{F}\,F\,\slashed{B}\,+\,i\,\frac{g}{2}\,\bar{F}\,
\sigma_i\,F \,\slashed{W^i}\,+\,i\,\frac{g_3}{2}\,\bar{F}\,\lambda_a\,F\,
\slashed{G^a} \;\;\; ,
\end{equation}
where $y_F$ is the field's hypercharge. If the field is an $SU(2)$ ($SU(3)$) 
singlet the $g$ ($g_3$) term is not present. The expression above is written for
the case $F$ is a triplet of $SU(3)$, for a field in the $\bar{3}$ 
representation we would have to replace $\lambda_a$ by $-\lambda_a^*$, as we
observed earlier for the scalar fields. Already we observe that the interactions
of the W's remain unchanged - these fields do not mix with others, so we obtain 
here the same vertices as in the SM (we have seen the W mass is now given by a 
different expression, but that does not affect the form of the vertices). 
Likewise the first seven gluons have the same gauge interactions they did 
previously - again they do not mix with any other field, and the fact that four 
of them now have mass does not affect their vertices. Where the differences with
the SM appear are in the interactions of the ``photon", ``$Z^0$" and the eighth 
gluon - for the case $F$ is a left lepton doublet $L\, =\, ({\nu_e}_L\;,\;e_L)$ 
(of hypercharge $1/2$; we wrote down that of the first lepton family, but this
calculation is valid for the remaining two) or a right singlet, $e_R$ (of 
hypercharge $-1$, if we are thinking of electrons), the lepton's gauge 
interactions will be
\begin{align}
{\nu_e}_L &: -\,\frac{1}{2}\,(\bar{\nu}_e\,\gamma^\mu\,\nu_e)_L\;(g^\prime\,
B_\mu \,-\,g\,W^3_\mu) \nonumber \\
e_L &: -\,\frac{1}{2}\,(\bar{e}\,\gamma^\mu\,e)_L\;(g^\prime\,B_\mu\,+\,g\,W^3_
\mu) \nonumber \\
e_R &: -\,g^\prime\,B_\mu \;\;\;\ .
\end{align}
Now, we can invert eq.~\eqref{eq:mix} so that
\begin{align}
B_\mu &= a_1\,\tilde{A}_\mu \,+\,a_2\,\tilde{Z}_\mu \,+\,a_3\,\tilde{G}_\mu 
\nonumber \\
W^3_\mu &= b_1\,\tilde{A}_\mu \,+\,b_2\,\tilde{Z}_\mu \,+\,b_3\,\tilde{G}_\mu
\nonumber \\
G^8_\mu &= c_1\,\tilde{A}_\mu \,+\,c_2\,\tilde{Z}_\mu \,+\,c_3\,\tilde{G}_\mu
\nonumber \;\;\; , 
\end{align}
and so the couplings of the fields $\tilde{A}$, $\tilde{Z}$, $\tilde{G}$ to 
leptons are given by
\begin{align}
{\nu_e}_L &: -\,\frac{1}{2}\,g\,a_i\,\left(\frac{g^\prime}{g}\,-\,
\frac{b_i}{a_i} \right) \nonumber \\
e_L &: -\,\frac{1}{2}\,g\,a_i\, \left(\frac{g^\prime}{g}\,+\,\frac{b_i}{a_i}
\right) \nonumber \\
e_R &: -\,g^\prime\,a_i \;\;\; .
\label{eq:cou}
\end{align}
For quarks the situation is more complex, we must take into account their colour
and the fact that the gauge fields will interact differently with them based
on it. So we have, for instance, for an up-type left quark (hypercharge 1/6),
\begin{equation} 
\bar{u}_L\,\slashed{\partial}\,u_L\;  \rightarrow \; \left(\frac{g^\prime}{6}\,
\slashed{B}\,+\,\frac{g}{2}\slashed{W^3}\right)\,(\bar{u}^1_L\,u^1_L+
\bar{u}^2_L\,u^2_L+\bar{u}^3_L\,u^3_L)\,+\,\frac{g_3}{2\sqrt{3}}\,(\bar{u}^1_L\,
u^1_L+\bar{u}^2_L\,u^2_L-2\,\bar{u}^3_L\,u^3_L)\,\slashed{G^8} \;\;\; .
\end{equation}
In the $g_3$ term we see that the gauge interactions with the quarks will depend
on their colours, even for the photon and $Z^0$! Thus, for the interactions 
between $\tilde{A}$, $\tilde{Z}$, $\tilde{G}$ and the quarks we will have
\begin{eqnarray}
u_L &:& g^\prime\,a_i\,\left[\frac{1}{6}\,+\,\frac{1}{2}\,\frac{b_i}{a_i}\,
\frac{g}{g^\prime}\,+\,\frac{g_3}{2\sqrt{3}g^\prime}\,\frac{c_i}{a_i}\,
\begin{pmatrix} 1 \\ 1 \\ -2 \end{pmatrix}\right] \nonumber \\
u_R &:& g^\prime\,a_i\,\left[\frac{2}{3}\,+\,\frac{g_3}{2\sqrt{3}g^\prime}\,
\frac{c_i}{a_i}\,\begin{pmatrix} 1 \\ 1 \\ -2 \end{pmatrix}\right] \nonumber \\
d_L &:& g^\prime\,a_i\,\left[\frac{1}{6}\,-\,\frac{1}{2}\,\frac{b_i}{a_i}\,
\frac{g}{g^\prime}\,+\,\frac{g_3}{2\sqrt{3}g^\prime}\,\frac{c_i}{a_i}\,
\begin{pmatrix} 1 \\ 1 \\ -2 \end{pmatrix}\right] \nonumber \\
d_R &:& g^\prime\,a_i\,\left[-\,\frac{1}{3}\,+\,\frac{g_3}{2\sqrt{3}g^\prime}\,
\frac{c_i}{a_i}\,\begin{pmatrix} 1 \\ 1 \\ -2 \end{pmatrix}\right] \;\;\; ,
\label{eq:couq}
\end{eqnarray}
where we used an obvious notation to indicate the $g_3$ terms (and them alone) 
act differently on the quarks, depending on their colours. It would seem all we
have done up to this point does not require supersymmetry at all - could we not
obtain the same results by adding scalar fields with the quantum numbers of 
$Q_L$ and $t_R$ to the SM? The answer is no as such a theory would have mixing 
terms between quarks and leptons. For instance, $t_R\,\tilde{Q}_L\,
\tilde{L}^c$ - the tilde denotes fermionic fields, $\tilde{L}$ is the lepton
fermionic doublet, $\tilde{L} \,=\,(\tilde{\nu}_L\,,\,\tilde{\tau}_L)$ (for the
third generation). This term involves the charge conjugate of $\tilde{L}$. It
does not occur in a supersymmetric theory because one cannot build SUSY 
lagrangeans using simultaneously a field and its charge conjugate. This is the 
same argument that implies the existence of a minimum of two Higgs doublets in 
the MSSM. Remember that the SUSY Yukawa term involving the leptonic superfield
$L$ - which contains $\tilde{L}$ - is $\lambda_\tau\,L\,H_1\,\tau_R$ (with 
$y_L = -1/2$, $y_{\tau_R} = +1$). 

Supersymmetry is thus necessary if we want to avoid mixing between quarks and
leptons. However, another type of mixing occurs, between the third colour 
component of the bottom quark and the charginos, and between the top quark, the
neutralinos and the eighth gluino. Such mixings are common when CCB occurs (see
for instance~\cite{tau} for mixing between $\chi^\pm$, $\chi^0$ and the leptonic
sector) and would in principle affect the electromagnetic interactions of the
third quark generation. At the very least such mixing could be used to impose 
constraints on the SUSY parameter space. We hope to address these questions in a
forthcoming paper~\cite{eu}. Another reason to work in the framework of 
supersymmetry is the well-known fact that adding scalars to the SM usually 
spoils asymptotic freedom, as the new fields have positive contributions to the 
$\beta$-function of the strong coupling constant. In a supersymmetric theory,
however, the field content is such that asymptotic freedom is preserved so that,
above the energy scale CCB occurs, the strong coupling runs with a 
supersymmetric $SU(3)$ $\beta$-function.  

\subsection{The photon}

Let us now apply these formulae to the massless field $\tilde{A}_\mu$, using the
values for $\{a_1\,,\,b_1\,,\,c_1\}$ given in eq.~\eqref{eq:coef}: because 
$b_1/a_1 \,=\,g^\prime/g$, we see from eqs.~\eqref{eq:cou} that $\tilde{A}$ does
{\em not} couple to the neutrinos. Plus, from the same expressions we see that 
it does couple identically with the left and right electrons. That is a further 
incentive to identify $\tilde{A}$ with the photon - the value of its coupling
with the electrons must therefore be the electric charge $e$, that is, $g^\prime
\,a_1\,=\,e$. With the results from~\eqref{eq:coef} we can therefore write
\begin{align}
a_1\;=\;\frac{g}{\sqrt{g^2 + {g^\prime}^2 + x^2}} & \;\;\;\; \mbox{, with  }
x\;=\;\frac{2}{\sqrt{3}}\,\frac{g\,g^\prime}{g_3} \;\;\; .
\label{eq:a1}
\end{align}
We see $a_1$ is a generalisation of the cosine of the Weinberg angle, involving
the $SU(3)$ coupling as well. As for the couplings of $\tilde{A}$ with the 
quarks we obtain, from eqs.~\eqref{eq:couq} and~\eqref{eq:coef},
\begin{align}
u_L\; , \; u_R &: \;\;\;\; \frac{1}{3}\,e\,\left[2 \,+\,\begin{pmatrix} 1 \\ 1 
\\ -2 \end{pmatrix} \right] \nonumber \\
d_L\; , \; d_R &: \;\;\;\; \frac{1}{3}\,e\,\left[-1 \,+\,\begin{pmatrix} 1 \\ 1 
\\ -2 \end{pmatrix} \right] \;\;\; .
\label{eq:phq}
\end{align}
If $\tilde{A}$ is indeed the photon then eq.~\eqref{eq:phq} is an expression of a 
very interesting phenomenon: the electric charge of the quarks depends on their
colour. In particular, for up-type quarks, the two first colours have charge 
$+1$ and the third is neutral. For down-type quarks the situation is reversed:
$d^{1,2}$ are chargeless and $d^3$ has charge $-1$. We can resume these results
in group theory terms, by noting that in this theory with broken colour and 
charge the electric charge operator is no longer $Q\,=\,T_3\,+\,Y$ but rather
\begin{equation}
Q\;=\; T_3\,+\,Y\,+\,C_8 \;\;\; ,
\label{eq:nQ}
\end{equation}
with $C_8$ the ``eighth colour", $C_8\,=\,\lambda_8/\sqrt{3}$. 
Notice how this new definition of the electric charge does not affect the charge
assignments of the leptons - they are colourless and thus the $\lambda_8$ term
does not come into play. Perhaps more interestingly, it does away with 
fractional electric charges altogether: the quarks' charges depend on their 
colours but are integer numbers. We recognise these as being identical to 
Han-Nambu quarks~\cite{han,pati} even though they are being obtained in a 
different context.

\subsection{The $Z^0$}

\begin{figure}[htb]
\epsfysize=9cm
\centerline{\epsfbox{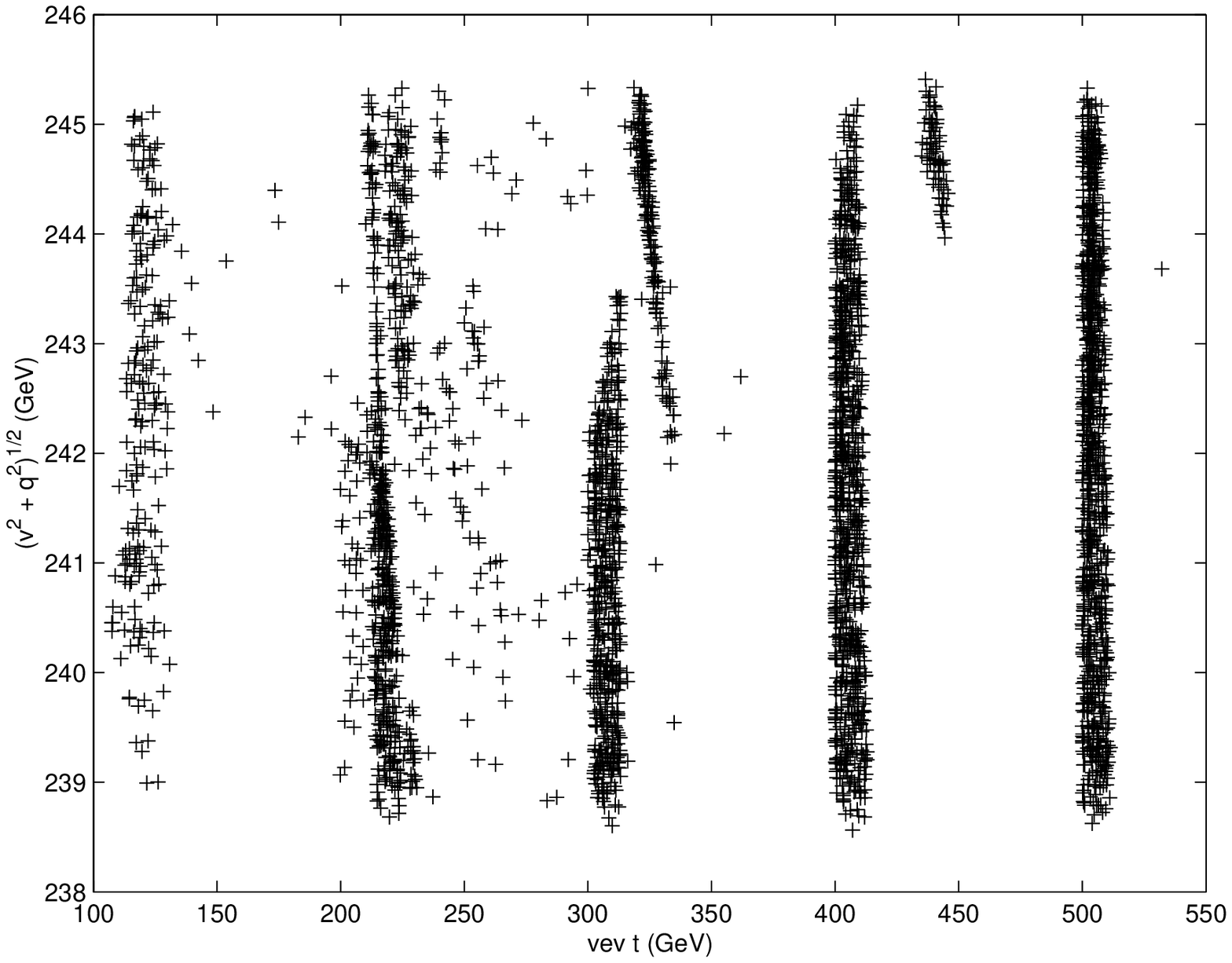}}
\caption{$\sqrt{v^2+q^2}$ {\em vs.} $t$, for the parameter space described in 
the text. Each of these points corresponds to a set of vevs $\{v, q, t\}$ for
whiche the $\tilde{Z}$ leptonic couplings are identical to those of the $Z^0$.}
\label{fig:vev}
\end{figure}

So far we have been able to ignore the values of the vevs in the theory, because
the coefficients $\{a_1, b_1, c_1\}$ do not depend on them. But trying to 
identify one of the two massive eigenstates as the $Z^0$ will require the
determination of $\{v, q, t\}$. There are two possibilities, the $Z^0$ can be 
the lighter or heavier of the two states. For reasons to be addressed in the 
next section it is more reasonable to assume $m_{\tilde{Z}} < m_{\tilde{G}}$ and
that is the case we discuss. The results for the other possibility are not 
qualitatively different from those presented here. From eqs.~\eqref{eq:cou} we
can write the couplings of $\tilde{Z}$ to the leptons as
\begin{align}
g_{\tilde{Z}_\nu} &= \frac{1}{2}\,(g\,b_2\,-\,g^\prime\,a_2) \nonumber \\
g_{\tilde{Z}_V} &= -\,\frac{1}{4}\,(g\,b_2\,+\,3\,g^\prime\,a_2) \nonumber \\
g_{\tilde{Z}_A} &= \frac{1}{4}\,(g\,b_2\,-\,g^\prime\,a_2) \;\;\; ,
\end{align}
where we have split the coupling of the $\tilde{Z}$ to charged leptons in an 
axial coupling (the terms that multiply the $\gamma^5$ matrix) and a vector 
coupling, as is usual. We immediately see that $g_{\tilde{Z}_A} \,=\,
g_{\tilde{Z}_\nu}/2$. In the SM at tree level we have an identical expression, 
$g^{SM}_{Z_A}\,=\,g^{SM}_{Z_\nu}/2$. In terms of the unit electric charge
$e$, the $W$ and $Z$ masses $M_W$, $M_Z$, we have
\begin{align}
g^{SM}_{Z_\nu} &= \frac{e\,M_Z^2}{2\,M_W\,\sqrt{M_Z^2-M_W^2}} \nonumber \\
g^{SM}_{Z_V} &= \frac{e\,(3\,M_Z^2-4\,M_W^2)}{4\,M_W\,\sqrt{M_Z^2-M_W^2}}
\;\;\; .
\end{align}
So the $\tilde{Z}$ has leptonic couplings identical to those of the $Z^0$ 
provided we have
\begin{align}
a_2 &= -\,\frac{e}{g^\prime}\,\frac{\sqrt{M_Z^2-M_W^2}}{M_W} \nonumber \\
b_2 &= \frac{e}{g}\,\frac{M_W}{\sqrt{M_Z^2-M_W^2}} \;\;\; .
\label{eq:lep}
\end{align}
In the SM these formulae reduce to the known results $-\sin\theta_W$, $\cos
\theta_W$ - here, due to eq.~\eqref{eq:a1} it is no longer true that $a_2^2 +
b_2^2 = 1$. At this point we can determine $\{v, q, t\}$ by using as inputs the
experimental values of $\{e, g_3, M_W, M_Z\}$. For given values of the vevs $g$
is given by $g^2 = 4\,M_W^2/(v^2 + q^2)$, $g^\prime$ is determined by 
eq.~\eqref{eq:a1} and from eq.~\eqref{eq:m2} we calculate $m_{\tilde{Z}}$. 
Following a simple minimisation procedure we can choose the vevs so that 
the leptonic couplings of the $\tilde{Z}$ are identical to those of the $Z$, 
requiring conditions~\eqref{eq:lep}. This procedure~\footnote{Doing this we are 
effectively equaling our tree-level predictions for the $\tilde{Z}$ couplings to
the tree-level results of the SM.} gives values of $m_{\tilde{Z}}$ that, in 
the vast majority of cases, verify $|m_{\tilde{Z}} - M_Z| < 10^{-3}$ GeV. The 
minimisation process is dependent on the initial guess for the $t$ vev - for the
same values of $\{e, g_3, M_W, M_Z\}$ different solutions $\{v, q, t\}$ may be
found. However, they all share a common trait, the fact that $\sqrt{v^2+q^2} 
\simeq 246$ GeV, as may be seen in fig.~\eqref{fig:vev}. The points in this plot
correspond to solutions obtained with different initial guesses for $t$ (from 
100 to 500 GeV in steps of 100), with random values of $\{e, g_3, M_W, M_Z\}$ 
taken in the intervals $0.3093 \leq e \leq 0.3173$, $1.063 \leq g_3 \leq 
1.1757$, $80.39 \leq M_W \leq 80.47$ GeV, $91.18 \leq M_Z \leq 91.20$ GeV. These
intervals correspond to the current experimental values and uncertainties for
these quantities (at the scale $M_Z$ for the couplings). Usually the larger the 
initial guess for $t$ the larger the number of solutions found (for the largest 
value tried, $t^{\mbox{\small{initial}}} = 2000$ GeV, about 87\% of the 
parameter space considered yielded a solution. We emphasise this is {\em not} an
exhaustive study of all possibilities involved, but the parameter space we 
considered is realistic and the solutions found, credible. It should also be 
noted that the difference between the quantity plotted in fig.~\eqref{fig:vev}, 
$\sqrt{v^2+ q^2}$, and the SM vev, 
\begin{equation}
v^{SM}\; =\; \frac{2\, M_W}{e\,M_Z}\,\sqrt{M_Z^2\,-\,M_W^2} \;\;\; ,
\end{equation} 
is smaller than $10^{-3}$ GeV for all the points considered. This is 
reassuring since this quantity is experimentally fixed by the lifetime of the 
muon (at tree level at least). This also implies the coupling $g$ is the same in
our theory and in the SM (not so for $g^\prime$, though). 

Having chosen $\{v, q, t\}$ to ``fix" the leptonic couplings we now look at the 
couplings of $\tilde{Z}$ to quarks. In the SM we have
\begin{align}
g^{SM}_{Z_{V_u}} &= \;\frac{e}{12}\,\frac{8 M_W^2\,-\,5 M_Z^2}{M_W\,\sqrt{M_Z^2
\,-\, M_W^2}} & g^{SM}_{Z_{A_u}} &= \; -\,\frac{e}{12}\,\frac{M_Z^2}{M_W\,
\sqrt{M_Z^2\,- \,M_W^2}} & \nonumber \\
g^{SM}_{Z_{V_d}} &= \; -\,\frac{e}{12}\,\frac{4 M_W^2\,-\,M_Z^2}{M_W\,\sqrt{
M_Z^2\,-\, M_W^2}} & g^{SM}_{Z_{A_d}} &= \; -\,g^{SM}_{Z_{A_u}}& \;\;\; .
\end{align}
Using eqs.~\eqref{eq:couq} and the fact the coefficients $\{a_2 , b_2\}$ are now
given by the expressions~\eqref{eq:lep} we obtain, for the axial couplings, 
$g_{\tilde{Z}_{A_u}} = g^{SM}_{Z_{A_u}}$ and $g_{\tilde{Z}_{A_d}} = 
g^{SM}_{Z_{A_d}}$. For the vector couplings the results is not as simple since 
there is dependence on the colour $i$ of the quark in question, 
\begin{align}
g_{\tilde{Z}_{V_u^i}} &= \; g^{SM}_{Z_{V_u}}\; + \;\frac{g_3}{2\sqrt{3}}\,c_2\,
\begin{pmatrix} 1 \\ 1 \\ -2 \end{pmatrix} \nonumber \\
g_{\tilde{Z}_{V_d^i}} &= \; g^{SM}_{Z_{V_d}} \; + \;\frac{g_3}{2\sqrt{3}}\,c_2\,
\begin{pmatrix} 1 \\ 1 \\ -2 \end{pmatrix} \;\;\; .
\label{eq:zq}
\end{align}

\subsection{The gluons}
\label{sec:glu}

\begin{figure}[htb]
\epsfysize=9cm
\centerline{\epsfbox{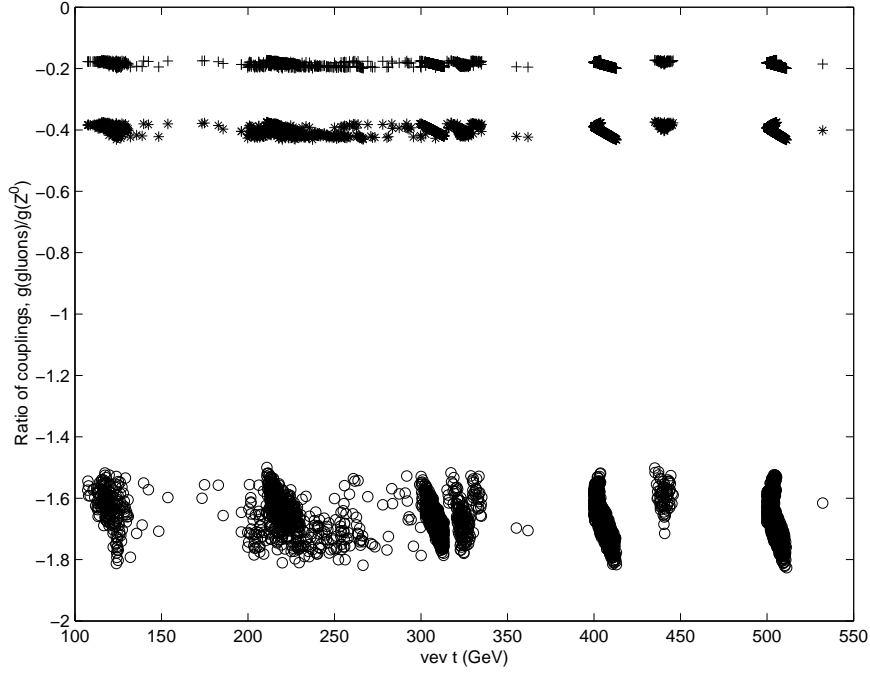}}
\caption{Ratio of $\tilde{G}$ gluon leptonic couplings to $Z^0$ ones - 
$g_{\tilde{G}_\nu}/g^{SM}_{Z_\nu}$ (crosses), $g_{\tilde{G}_A}/g^{SM}_{Z_A}$ 
(stars), $-g_{\tilde{G}_V}/g^{SM}_{Z_V}$ (circles) - notice the minus sign in 
this last ratio.}
\label{fig:glg}
\end{figure}

Five gluons are massive in this theory. One of them, which we call $\tilde{G}$, 
due to its mixing with $\tilde{\gamma}$ and $\tilde{Z}$ couples directly to the
leptons. First of all we must ask what are the masses of these gluons - as 
explained earlier we have two possibilities, $m_{\tilde{G}} < m_{\tilde{Z}}$ and
$m_{\tilde{G}} > m_{\tilde{Z}}$. Since $\tilde{G}$ could in principle be 
produced in $e^+e^-$ collisions we assume that the first possibility is ruled 
out by experiment. In fact, given LEP2's results and operational energy we must
probably require $m_{\tilde{G}} > 200$ GeV, and this not counting possible 
interference effects~\footnote{Remember that the $Z^0$ makes its presence 
``felt" in $e^+e^-$ collisions even below its peak.}. For the moment let us not 
worry about absolute experimental bounds and simply consider the case 
$m_{\tilde{G}}> m_{\tilde{Z}}$. As we showed in the last section, requiring this
model to correctly reproduce the SM's tree-level predictions for electroweak 
interactions constrains $v^2+q^2 \simeq (246 \mbox{ GeV})^2$. It is the value 
of the $t$ vev that thus will set the scale for the mass of the gluons. 
Therefore the fact that different sets of solutions $\{v, q, t\}$ may be 
obtained, for the same input values of $\{e, g_3, M_W, M_Z\}$, for different 
initial guesses for t, gains added importance. In fig.~\eqref{fig:mgl} we plot 
the masses of $\tilde{G}$ and the remaining four massive gluons - call them $G$,
and remember they are degenerate in mass - against $t$, for several different 
initial guesses. We always have $m_G < m_{\tilde{G}}$ and observe that $m_G >
76$ GeV, a fact that will prove experimentally relevant in a short while. An 
initial guess of 2000 GeV did produce acceptable results and larger values still
could have been used, yielding even larger gluon masses. Our procedure is not 
able to make definite predictions for the gluon masses, as we find no upper 
bound for them. We should be able to restrict the values of $t$ by analysing 
their impact in quark masses~\cite{eu}. Using eqs.~\eqref{eq:cou}, 
\eqref{eq:couq} just like we did for the $\tilde{Z}$ we can find the couplings 
of $\tilde{G}$ to the fermions. In fig.~\eqref{fig:glg} we plot the ratio of the
neutrino, axial electron and vector electron couplings for $\tilde{G}$ and $Z^0$
couplings. We see that the $\tilde{G}$ couplings are of the same order of 
magnitude than the $Z^0$ ones, and mostly independent of either the input values
$\{e, g_3, M_W, M_Z\}$ and the $t$ vev - the difference in signs is due to the 
orthogonality of the coefficients $\{a_i, b_i, c_i\}$. 

\begin{figure}[htb]
\epsfysize=9cm
\centerline{\epsfbox{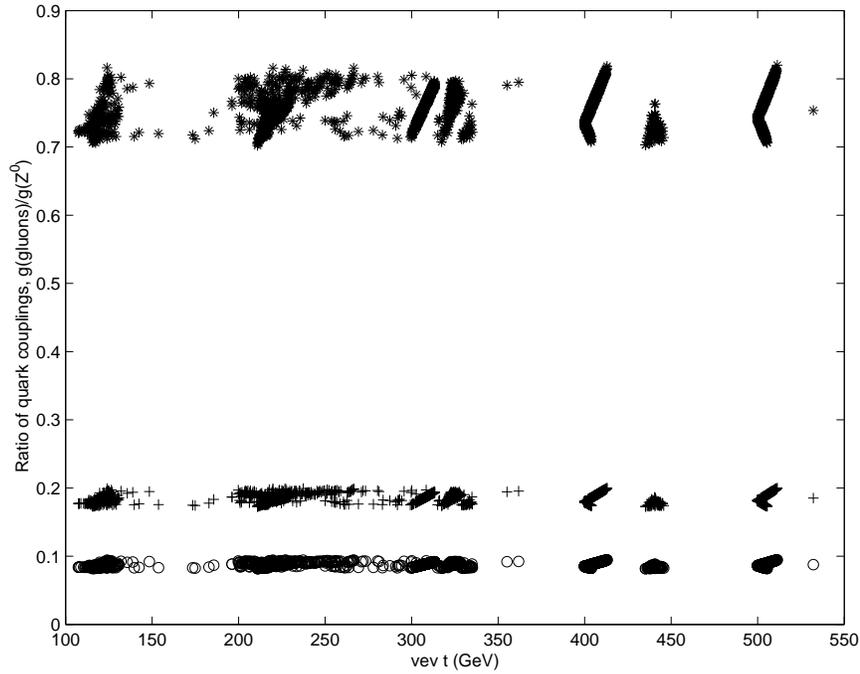}}
\caption{Ratio of $\tilde{G}$ gluon quark couplings to $Z^0$ ones - 
$-g_{\tilde{G}_{A_q}}/g^{SM}_{Z_{A_q}}$ (crosses - notice the minus sign), 
$g_{\tilde{G}_{V_u}}/g^{SM}_{Z_{V_u}}$ (stars), $g_{\tilde{G}_{V_d}}/g^{SM}_{Z_{
V_d}}$ (circles).}
\label{fig:gqg}
\end{figure}

\begin{figure}[htb]
\epsfysize=9cm
\centerline{\epsfbox{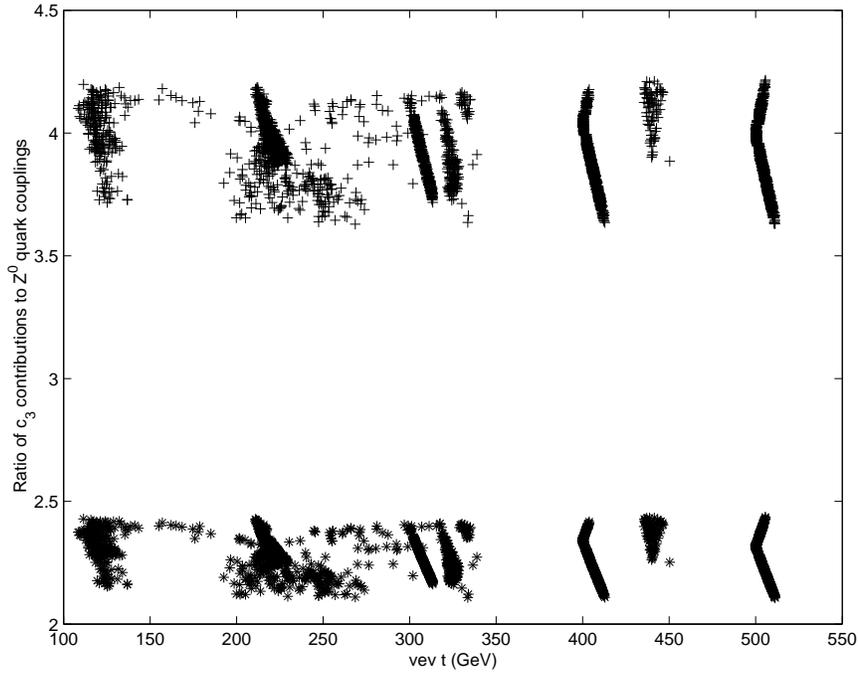}}
\caption{Ratio of $g_3\,c_3/2\,\sqrt{3}$ to $-g^{SM}_{Z_{V_u}}$ (crosses - 
notice the minus sign) and $g^{SM}_{Z_{V_d}}$ (stars).}
\label{fig:conc3}
\end{figure}

Finally, in fig.~\ref{fig:gqg} we plot the ratio of the quark couplings of 
$\tilde{G}$ and $Z^0$. As before we see there is little variation in the 
strength of the couplings with the value of either $t$ or the input values we 
considered. In this figure we do not consider the terms proportional to $c_3$ in
eq.~\eqref{eq:couq}, those who distinguish between the quarks' colour - if this 
gluon is not confined then we would expect that (see section~\ref{sec:test}) the
$c_3$ contributions will sum to zero in any $q\bar{q}$ reaction; if the gluon is
confined then those contributions will be of interest. In fig.~\eqref{fig:conc3}
we plot the ratio of $g_3\,c_3/2\,\sqrt{3}$ to the $Z$ vector couplings of the 
$u$ and $d$ quarks and see, without surprise, that these ``pure colour" 
couplings are much stronger than the ``pure electroweak" ones we plotted in 
fig.~\ref{fig:gqg}~\footnote{There are no $c_3$ contributions to the axial 
couplings.}. 

\subsection{Gauge boson self-couplings}
\label{sec:gau}

In a non abelian theory gauge bosons can interact between themselves due to the
presence in the lagrangean of cubic and quartic terms of the form
\begin{equation}
-\,g\,f_{abc}\,\left(\partial_\mu A_\nu^a\right)\,{A^b}^\mu\,{A^c}^\nu\;-\;
\frac{g^2}{4}f_{abc}\,f_{ade}\,A^b_\mu\,A^c_\nu\,{A^d}^\mu\,{A^e}^\nu
\end{equation}
for a gauge group with structure constants $f_{abc}$ and a gauge coupling $g$.
The Feynman rules for the vertices corresponding to these terms are
\begin{align}
\mbox{Cubic:} &  \hspace{0.6cm} g\, f_{abc}\left[(r-q)_\lambda\, g_{\mu\nu}\,+\,
(q-p)_\nu\, g_{\lambda\nu}\,+\,(p-r)_\mu\, g_{\nu\lambda}\right] \nonumber \\
\mbox{Quartic:} & \;\;\; -\,i\,g^2\,\left[f_{abc}\,f_{ade}\,(g_{\lambda\nu}\,
g_{\mu\rho}\,-\,g_{\mu\nu}\,g_{\lambda\rho})\,+\,f_{adc}\,f_{abe}\,(g_{\lambda
\nu}\,g_{\mu\rho}\,-\,g_{\mu\lambda}\,g_{\nu\rho})\,+ \right. \nonumber \\
 &  \hspace{1.5cm} \left. f_{abd}\,f_{ace}\,(g_{\lambda\mu}\,g_{\nu\rho}\,-\,
g_{\mu\nu}\,g_{\lambda\rho})\right]\;\;\; ,
\label{eq:ver}
\end{align}
where $\{p, q, r\}$ are the 4-momenta of each particle in the vertex and the
greek letters their Lorentz indices. In the SM the mixing between $B_\mu$ and 
$W^3_\mu$ produces such vertices as $\gamma\,W^+\,W^-$, $Z^0\,W^+\,W^-$, $\gamma
\,\gamma\,W^+\,W^-$, $\gamma\,Z^0\,W^+\,W^-$, etc. There are however no vertices
mixing $SU(2)\times U(1)$ gauge fields with $SU(3)$ ones. In our theory, with 
broken colour symmetry, the mixing prescribed by eq.~\eqref{eq:mix} changes that
situation - every time we consider terms involving the fields $W^3_\mu$ or 
$G^8_\mu$ in eq.~\eqref{eq:ver} vertices involving both gluons and electroweak 
gauge bosons appear. Their respective couplings are given by the product of
$g$ ($g_3$) and the coefficients $b_i$ ($c_i$). Let us observe however that once
again our theory perfectly reproduces the electroweak SM results - having chosen
the vevs so that eq.~\eqref{eq:lep} is satisfied we conclude that the 
$\tilde{Z}$ coupling in any gauge boson vertex is
\begin{equation}
g\,b_2\; = \; e\,\frac{M_W}{\sqrt{M_Z^2-M_W^2}} \; = \; g\,(\cos\theta_W)^{SM}
\end{equation}
exactly the same as the analogous $Z^0$ coupling~\footnote{We emphasise that
$(\cos\theta_W)^{SM}$ is a SM quantity, it is {\em not} given, in our theory, by
$g/\sqrt{g^2+{g^\prime}^2}$, due to eq.~\eqref{eq:a1}.}. Likewise for 
$\tilde{A}$, the coupling will be $g\,b_1\,=\,g^\prime\,a_1\,=\,e$, just like 
for the photon. Thus the cubic and quartic vertices involving only the 
electroweak gauge bosons remain unchanged. We now list all the new 
self-interaction gauge boson vertices and their couplings (in listing the 
quartic couplings we leave out a factor of $-i$). 

From $SU(2)$ we find a single new cubic vertex, $\tilde{G}\,W^+\,W^-$, with 
coupling obviously given by $g\,b_3$. As for quartic vertices, with two $W^3$ 
fields in the quartic terms of~\eqref{eq:ver} we obtain three vertices with 
$\tilde{G}$ fields, 
\begin{align}
\tilde{G}\,\tilde{G}\,W^+\,W^- \;\;\;\;\; \Longleftrightarrow & \;\;\;\;\; 
(g\,b_3)^2 \nonumber \\
\tilde{G}\,\tilde{\gamma}\,W^+\,W^- \;\;\;\;\; \Longleftrightarrow & \;\;\;\;\;
e \,g\,b_3 \nonumber \\
\tilde{G}\,\tilde{Z}\,W^+\,W^-  \;\;\;\;\; \Longleftrightarrow & \;\;\;\;\; -\,
g^2\,b_3 \,(\cos\theta_W)^{SM} \;\;\; .
\end{align}

For $SU(3)$ self-couplings, the group's structure constants are well known, 
their non-zero values given by
\begin{align}
&f_{123}\;=\;1 \nonumber \\
&f_{147}\;=\;f_{165}\;=\;f_{246}\;=\;f_{257}\;=\;f_{345}\;=\;f_{376}\;=\;
\frac{1}{2} \nonumber \\
&f_{845}\;=\;f_{867}\;=\;\frac{\sqrt{3}}{2} \;\;\; ,
\label{eq:str}
\end{align}
and all permutations of these indices bearing in mind that $f_{abc}$ is entirely
antisymmetric. Notice that $G^8$ (and thus $\tilde{\gamma}$, $\tilde{Z}$, 
$\tilde{G}$) does {\em not} couple directly to the massless gluons (there are no
non-zero $f_{81i}$, $f_{82i}$ or $f_{83i}$ structure constants). The main 
difference 
between the SM cubic gluon vertices and our own is the fact five gluons are now 
massive. However, for vertices not involving $G^8$ the vertices themselves are 
unchanged (the propagators for $G^{4\ldots 7}$, of course, are now those of 
massive spin-1 particles). Three new vertices appear,  
\begin{align}
\tilde{G}\,G\,G \;\;\;\;\; \Longleftrightarrow & \;\;\;\;\; \frac{\sqrt{3}}{2}\,
g_3 \,c_3 \nonumber \\
\tilde{\gamma}\,G\,G \;\;\;\;\; \Longleftrightarrow & \;\;\;\;\; 
\frac{\sqrt{3}}{2} \,g_3 \,c_1 \;\;=\;\;e \nonumber \\
\tilde{Z}\,G\,G \;\;\;\;\; \Longleftrightarrow & \;\;\;\;\;\frac{\sqrt{3}}{2}\,
\,g_3 \,c_2 \;\;\; .
\end{align}
Notice that the $\tilde{\gamma}\,G\,G$ vertex implies the $4 \ldots 7$ gluons
have electric charge in this theory - the interaction with $\tilde{\gamma}$
changes the gluons' colour as well, though, in agreement with the 
formula~\eqref{eq:nQ} for the electric charge, now containing a colour 
contribution. From this point forward we will name the massive $4, \ldots 7$ 
gluons by $G_1^\pm$, $G_2^\pm$ and the massless $1, \ldots 3$ ones by $G$. Also notice 
that, with $M_G > 76$ GeV as we found in fig.~\eqref{fig:mgl}, the decay 
$\tilde{Z} \rightarrow G_i^+\,G_i^-$ is kinematically forbidden - again, the 
existing experimental data could be used to find lower bounds on $M_G$. We must 
emphasise that because $G^8$ does not couple directly to massless gluons there 
are no vertices of the form $\tilde{\gamma}\,G\,G$ or $\tilde{Z}\,G\,G$, 
interactions that would certainly have been observed already - an extra argument
to support the identification $\tilde{\gamma} \equiv \gamma$ and $\tilde{Z} 
\equiv Z^0$. 

Given the structure constants~\eqref{eq:str} we may have quartic vertices with 
one $G^8$ field - these will necessarily include a massless gluon $G$ and two 
charged gluons. We therefore have three new vertices, 
\begin{align}
\tilde{\gamma}\,G\,G_i^+\,G_i^- \;\;\;\;\; \Longleftrightarrow & \;\;\;\;\; 
\frac{\sqrt{3}}{4}\,g_3^2\,c_1 \;\;=\;\;\frac{g_3}{2}\,e \nonumber \\
\tilde{Z}\,G\,G_i^+\,G_i^- \;\;\;\;\; \Longleftrightarrow & \;\;\;\;\; 
\frac{\sqrt{3}}{4}\,g_3^2\,c_2 \nonumber \\
\tilde{G}\,G\,G_i^+\,G_i^- \;\;\;\;\; \Longleftrightarrow & \;\;\;\;\;
\frac{\sqrt{3}}{4}\,g_3^2\,c_3 \;\;\; .
\end{align}
Finally, the vertices with two $G^8$ ``legs" produce 
\begin{align}
\tilde{\gamma}\,\tilde{\gamma}\,G_i^+\,G_i^- \;\;\;\;\; \Longleftrightarrow & 
\;\;\;\;\; \frac{3}{4}\,g_3^2\,c_1^2\;\;=\;\;\frac{1}{2}\,e^2 
\nonumber \\
\tilde{\gamma}\,\tilde{Z}\,G_i^+\,G_i^- \;\;\;\;\; \Longleftrightarrow & \;\;\;\;\; 
\frac{3}{4}\,g_3^2\,c_1 \,c_3 \;\;=\;\;\frac{\sqrt{3}}{2}\,g_3\,c_2\, e 
\nonumber \\
\tilde{Z}\,\tilde{Z}\,G_i^+\,G_i^- \;\;\;\;\; \Longleftrightarrow & \;\;\;\;\;
\frac{3}{4}\,g_3^2\,c_2^2 \nonumber \\
\tilde{Z}\,\tilde{G}\,G_i^+\,G_i^- \;\;\;\;\; \Longleftrightarrow & \;\;\;\;\;
\frac{3}{4}\,g_3^2\,c_2\,c_3 \nonumber \\
\tilde{G}\,\tilde{G}\,G_i^+\,G_i^- \;\;\;\;\; \Longleftrightarrow & \;\;\;\;\;
\frac{3}{4}\,g_3^2\,c_3^2 \;\;\; .
\end{align}
Throughout this section we showed our broken QCD theory predicts the existence
of vertices inexistent in the SM - this, like the existence of massive gluons 
and integer quark charges, would be an unequivocal way of distinguishing both 
models.

\section{Testing the theory}
\label{sec:test}

In this theory the leptons have the same charges and couplings they do in the
SM, for sensible choices of the vevs involved. For the most their electroweak 
gauge interactions are thus identical in both theories. The only exceptions are 
a new interaction in the broken $SU(3)$ theory - the vertex $e^+\,e^-\,
\tilde{G}$, already discussed - and two-photon physics. Other potential 
differences are found in the strong sector. 

\subsection{Hadrons}
\label{sec:had}

What one observes in experiments are not individual quarks of a given colour but
rather mesons and baryons, which are singlets of $SU(3)$. In this theory the 
photon itself carries colour degrees of freedom, hence having colour cannot be 
synonymous of being confined. Nevertheless we continue to assume that bound 
states of the strong interactions are $SU(3)$ colour singlets~\footnote{I thank 
Tim Jones and Carlo Becchi for discussions on this matter.}. In our case we have
broken $SU(3)$ to a residual symmetry $SU(2)$ - if the unbroken gauge group is 
$SU(2)$ how can we require the hadronic bound states be singlets of $SU(3)$? 
After all, confinement can be interpreted as being due to a quark-quark 
potential that grows linearly with the quark separation $r$, $V_{qq}\,=\,\alpha
\,r$. We argue that this is a consequence of the non-linearity of the non 
abelian field equations, rather than the masslessness of the gauge boson 
carriers. We also remark that confinement is a non-perturbative process of which
little is yet known, and that
no clear demonstration exists that an unbroken $SU(3)$ gauge group leads to 
quarks confined as singlets of that same $SU(3)$. In light of that we would 
argue that our hypothesis - the unbroken gauge group is $SU(2)$ but the quark
bound states are singlets of $SU(3)$ - is as much a leap in the dark as the SM
one - unbroken $SU(3)$ implies confinement of quarks in $SU(3)$ singlet states.
It is also worth mentioning the work by Bailin and Love~\cite{bai} proposing 
that the colour symmetry be $SU(2)$, not $SU(3)$. Their basic argument was that
baryons are states with wave functions of the form $\epsilon_{ijk}\,\psi^i\,
\phi^j\,\chi^k$ and $\epsilon_{ijk}$ is an $SU(2)$ tensor; this structure can
therefore constitute an $SU(2)$ singlet as long as the quarks $\psi$, $\phi$, 
$\chi$ and in the triplet representation of that group. This result is 
interesting to us because to reproduce the SM results from our ICQ theory all we
really need is for the colour wave functions of the hadrons to have the same
structure (see below). In the framework of Bailin and Love that would occur and 
this would fit quite well with our unbroken $SU(2)$ symmetry. Their proposal was
not successful because two-loop corrections destroyed any possibility it had of 
predicting asymptotic freedom for the strong coupling constant~\cite{bai2}. As 
was already explained that would not be a problem in our case, since above the
CCB scale the running of $\alpha_S$ would be the ``normal" supersymmetric 
$SU(3)$ one. However the fact remains that we would need our quarks to be in the
triplet representation of $SU(2)$, when in fact they are from the start triplets
of $SU(3)$. 

With this hadronization hypothesis we find that the new formula for the electric
charge, eq.~\eqref{eq:nQ}, even though it changes the charge assignments for 
each quark, does not alter the overall electric charge of hadrons. For baryons, 
summing up the charge of the three quarks will correspond, each of their colours
being different, to taking the trace of $\lambda_8$; the end result is thus the 
SM one. For mesons each quark of colour $i$ comes with an antiquark of opposite 
colour and so the $\lambda_8$ contributions in eq.~\eqref{eq:nQ} are cancelled 
once more. But what about the distribution of these electric charges inside the 
hadrons? Let us consider the example of the proton, composed of two up quarks 
and a down one.  Because in our model the electric charge varies with the 
quarks' colour we must consider the proton's colour wave function~\footnote{For 
the purpose of the electric charge we need not worry about the spin wave 
function.},
\begin{equation}
\psi_{colour}^p \;=\; \frac{1}{\sqrt{6}}\,\left[u^1 u^2 d^3\,-\,u^2 u^1 d^3\,+\,
u^2 u^3 d^1\,-\,u^3 u^2 d^1\,+\,u^3 u^1 d^2\,-\,u^1 u^3 d^2 \right]\;\;\; .
\label{eq:psicp}
\end{equation}
From section~\eqref{sec:int} we know that $q(u^1) = q(u^2) = - q(d^3) = +1$ (in
units of $e$) and $q(d^1) = q(d^2) = q(u^3) = 0$. We thus see that in the wave 
function above the first two terms correspond to a system of three unit electric
charges, two positive and one negative. In the four remaining terms there
is a single electric charge - of the quarks $u^1$ or $u^2$ - making up for the
proton's overall charge. This means for instance that, according to our model,
in a deep elastic/inelastic scattering experiment an approaching electron would,
in an electromagnetic interaction, ``see" a proton as a three-charge system with
probability $1/3$ and as an electric monopole with probability $2/3$. If we 
define the ``proton charge wave function", it will be given by a sum of two 
pieces, one describing a monopole and another a triplet of charges, 
\begin{equation}
\psi_{charge}^p \;=\; \frac{1}{\sqrt{3}}\,\left[\psi_p^{(++-)}\,+\,\sqrt{2}\,
\psi_p^{(+)}\right] \;\;\; .
\end{equation}
This of course corresponds to the definition
\begin{equation}
\psi_p^{(++-)} \;=\; \frac{1}{\sqrt{6}}\,\left[u^1 u^2 d^3\,-\,u^2 u^1 d^3
\right]
\end{equation}
and a similar definition for $\psi_p^{(+)}$ involving the four remaining terms
of eq.~\eqref{eq:psicp}. The overall proton charge thus remains the same but 
while in the SM that charge is distributed in three ``lumps", in our model we 
find it this the superposition of two different states. Of course we are looking
at the proton in an extremely simplistic manner - not taking into account the 
sea quarks, for instance. This result seems quite remarkable though, and, for 
regions of transferred momentum not too high, it could be relevant. The same 
reasoning may be applied to the neutron, composed of two $d$ quarks and a $u$ 
one. Once again studying the colour wave function we obtain
\begin{equation}
\psi_{charge}^n \;=\; \frac{1}{\sqrt{3}}\,\left[\psi_n^{(0)}\,+\,\sqrt{2}\,
\psi_n^{(+-)}\right] \;\;\; .
\label{eq:qneu}
\end{equation}
That is to say, in a deep neutron scattering experiment the electromagnetic 
interaction should ``see" the neutron as an electric dipole with probability
$2/3$ and a state composed of only neutral charges with probability $1/3$. These
neutral charges correspond to the terms $(u^3 d^1 d^2\,-\,u^3 d^2 d^1)/\sqrt{6}$
in the neutron colour wave function, where all the quarks involved are neutral. 
The striking thing about eq.~\eqref{eq:qneu} is that - in this simplistic 
approach, at least - it predicts the neutron is {\em never} a state with three 
charges! We can follow this line of reasoning to the $\pi^0$ meson, described by
the colour wave function
\begin{equation}
\psi^{\pi^0}_{colour} \;=\; \frac{1}{\sqrt{6}}\,\left(u^1 \bar{u}^1\,+\,u^2 
\bar{u}^2\,+\,u^3 \bar{u}^3\,-\,d^1 \bar{d}^1\,-\,d^2 \bar{d}^2\,-\,d^3 
\bar{d}^3 \right)\;\;\; ,
\label{eq:pi0}
\end{equation}
to obtain
\begin{equation}
\psi_{charge}^{\pi^0} \;=\; \frac{1}{\sqrt{2}}\,\left[\psi_{\pi^0}^{(0)}\,+\,
\psi_{\pi^0}^{(+-)}\right] \;\;\; ,
\end{equation}
which indicates the $\pi^0$ should be observed as an electric dipole with 
probability $1/2$ - not $1$ as it would be in the SM. The most interesting case,
though, is that of the $\pi^+$ ($\pi^-$). Its colour wave function being
\begin{equation}
\psi^{\pi^+}_{colour} \;=\; \frac{1}{\sqrt{3}}\,\left(u^1 \bar{d}^1\,+\,u^2 
\bar{d}^2\,+\,u^3 \bar{d}^3\right)\;\;\; ,
\end{equation}
we see that in each term there is a single quark with charge - meaning, the 
charge wave function will be
\begin{equation}
\psi_{charge}^{\pi^+} \;=\; \psi_{\pi^+}^{(+)} \;\;\; ,
\end{equation}
and the $\pi^+$ (likewise the $\pi^-$) should, with $100\%$ probability, be 
observed as an electric {\em monopole}, never as a {\em dipole}. These are 
fascinating predictions that, one hopes, may be used to test the validity of 
this theory. 

One might expect the new quark electric charge assignments to have some impact 
in the magnetic momenta of baryons but surprisingly that is not the case. 
Consider, for instance, the proton. For the calculation of its magnetic momentum
it is necessary to take into account the spin part of the wave function, so we 
describe the proton by the (spin up) state
\begin{equation}
|p\uparrow >\;=\;\frac{1}{\sqrt{18}} \left[u\uparrow u\downarrow d\uparrow\,+\,
u\downarrow u\uparrow  d\uparrow\,-\,2\,u\uparrow u\uparrow d\downarrow\,+\, 
\mbox{permutations} \right]\, \otimes \, \psi_{colour}
\label{eq:psip}
\end{equation}
with
\begin{equation}
\psi_{colour} \;=\; \frac{1}{\sqrt{6}} \left[1\,2\,3\;-\;2\,1\,3\;+\;
2\,3\,1\;-\;3\,2\,1\;+\;3\,1\,2\;-\;1\,3\,2\right]
\end{equation}
in the usual notation. The magnetic momentum of the proton is thus given by
\begin{equation}
\mu_p \;=\;\sum_{i=quarks}\,<p\uparrow |\,\mu_i\,|p\uparrow > \;\;\; ,
\end{equation}
with $\mu_i$ the magnetic momentum of each quark, given by
\begin{equation}
\mu_u\;=\; q_u\,\hat{\mu}_u\,\sigma_u\hspace{0.5cm}, \hspace{0.5cm}i\mu_d\;=\; 
q_d\, \hat{\mu}_d\,\sigma_d \;\;\; ,
\end{equation}
$\sigma_u$ and $\sigma_d$ being the spin operators of the up and down quarks,  
$q_u$ and $q_d$ their respective charges (in units of $e$) and 
\begin{equation}
\hat{\mu}_u\;=\;\frac{e}{m_u}\hspace{0.5cm}, \hspace{0.5cm}\hat{\mu}_d\;=\;
\frac{e}{m_d} \;\;\; .
\end{equation}
In the SM the charge/spin are independent of colour so the colour wave function
need not concern us, $q_u = 2/3$, $q_d = -1/3$ and we obtain
\begin{align}
\mu_p^{SM} &=\;\frac{1}{18}\,3\,\left\{\frac{2}{3}\,\hat{\mu}_u\left(
\frac{1}{2}\,-\,\frac{1}{2}\right)\,-\,\frac{1}{3}\,\hat{\mu}_d\left(+
\frac{1}{2}\right)\,+\,\frac{2}{3}\,\hat{\mu}_u\left(-\frac{1}{2}\,+\,
\frac{1}{2}\right)\,-\,\frac{1}{3}\,\hat{\mu}_d\left(+\frac{1}{2}\right) 
\right. \nonumber \\
 & \left. \hspace{1.7cm}+\,4\,
\left[\frac{2}{3}\,\hat{\mu}_u\left(\frac{1}{2}\,+\,\frac{1}{2}\right)\,-\,
\frac{1}{3}\,\hat{\mu}_d\left(-\frac{1}{2}\right)\right] \right\} \nonumber \\
 &=\;\frac{1}{18}\,\left(8\,\hat{\mu}_u\,+\,\hat{\mu}_d \right)\;\;\; ,
\end{align}
the factor of $3$ taking care of the particle permutations in 
formula~\eqref{eq:psip}. In the broken QCD model we need to consider the effect
$\psi_{colour}$ has in the quarks' charge in each term. For instance, the
$u\uparrow u\downarrow d\uparrow$ term produces six terms in the wave function,
\begin{align}
u\uparrow u\downarrow d\uparrow &\longrightarrow \;\frac{1}{6\sqrt{3}}\,\left(
u^1\uparrow u^2\downarrow d^3\uparrow\,-\,u^2\uparrow u^1\downarrow d^3\uparrow
\,+\,u^2\uparrow u^3\downarrow d^1\uparrow\,-\,u^3\uparrow u^2\downarrow 
d^1\uparrow\,+ \right. \nonumber \\
 & \;\;\; \hspace{1.6cm} \left. u^3\uparrow u^1\downarrow d^2\uparrow\,-\,u^1
\uparrow u^3 \downarrow d^2\uparrow\right) \;\;\; .
\end{align}
Their contribution to the magnetic momentum is $-\hat{\mu}_d/108$. The 
$u\downarrow u\uparrow d\uparrow$ term has an identical contribution and the 
$2\,u\uparrow u\uparrow d\downarrow$ term, $(4\hat{\mu}_u\, +\,\hat{\mu}_d)/27$.
The particle permutations of eq.~\eqref{eq:psip} introduce a global 
multiplicative factor of 3 so that the final magnetic momentum is
\begin{equation}
\mu_p \;=\; 3\,\left[\frac{1}{27}\,(4\hat{\mu}_u\,+\,\hat{\mu}_d)\,-\,
\frac{1}{108}\,\hat{\mu}_d\,-\,\frac{1}{108}\,\hat{\mu}_d\right] \;=\; 
\frac{1}{18}\,\left(8\,\hat{\mu}_u\,+\,\hat{\mu}_d \right)\;\;\; ,
\end{equation}
exactly the SM result. It is possible to generalise this calculation  for the 
case of any baryon.

To close this section let us briefly consider the impact of the new quark charge
assignments on the momentum distribution functions for partons. The 
proton's structure functions $F_1$ and $F_2$ are given by
\begin{equation}
F_2(x)\;=\;x\,F_1(x)\;=\; x\sum_{i=quarks}Q_i^2\,f_i(x) \;\;\; ,
\end{equation}
$f_i(x)$ the probability distribution of finding quark $i$ of electric charge 
$Q_i$ inside the proton with a fraction $x$ of the total momentum. In a first 
approximation each quark $u$ carries, in average, as much momentum as 
the $d$ ones, 
\begin{equation}
\int_0^1\,x\,u(x)\,dx \;=\; \int_0^1\,x\,d(x)\,dx 
\label{eq:ud}
\end{equation}
and so in the SM we have
\begin{equation}
F_2(x)\;=\;x\,\left[2\,\left(\frac{2}{3}\right)^2 u(x)\,+\,\left(\frac{1}{3}
\right)^2 d(x)\right]
\label{eq:f2}
\end{equation}
so that~\footnote{This is not the usual notation used - the factor of $2$ 
affecting $u(x)$ in eq.~\eqref{eq:f2} is not present as one generally 
considers $u(x)$ as the probability distribution of {\em both} $u$ quarks. The
usual convention is to consider the $u$ quarks carry twice as much momentum as
the $d$ one. We adopt this non-standard notation for the simple reason it is 
easier to generalise to our theory.}, from eq.~\eqref{eq:ud}, 
\begin{equation}
\int_0^1\,F_2(x)dx \;=\; \int_0^1\,x\,d(x)\,dx \;\simeq\; 0.18 \;\;\; ,
\end{equation}
this value measured in deep inelastic scattering (DIS) experiments. So the 
momentum fraction carried by the $u$ and $d$ quarks is 
\begin{equation}
\int_0^1\,x\,\left[d(x)\,+\,2\,u(x)\right]\,dx \;\simeq\; 0.18\,+\,2\,.\,0.18 \;=\;
0.54 \;\;\; .
\label{eq:fm}
\end{equation}
Because DIS experiments probe only charged particles inside the proton this 
result is used as indirect evidence that about $46\%$ of the momentum of the 
proton is carried by neutral particles - the gluons. In our model we know 
several colour components of the quarks are electrically neutral, so, how are
these results affected? Due to the charge assignments we will have, instead of
eq.~\eqref{eq:f2}, 
\begin{equation}
F_2(x)\;=\;x\,\left[u^1(x)\,+\,u^2(x)\,+\,d^3(x)\right] \;\;\; ,
\end{equation}
$u^1(x)$, $u^2(x)$ ($d^3(x)$) the probability distributions for the $u$ ($d$) 
quarks of momentum fraction $x$ and colours 1 and 2 (3) respectively. Since we 
need to consider the distribution probabilities for each colour the hypothesis
equivalent to eq.~\eqref{eq:ud} is to assume the colour
component of each quark carries the same momentum in average, that is,
\begin{equation}
\int_0^1\,x\,u^i(x)\,dx \;=\; \int_0^1\,x\,d^i(x)\,dx
\end{equation} 
and so
\begin{equation}
\int_0^1\,F_2(x)dx \;=\; \displaystyle{\sum_{i=1}^3}\,\int_0^1\,x\,d^i(x)\,dx 
\;\;\; .
\end{equation}
Further assuming the $d$ quark distribution is independent of the colour index
we will have $d^i(x) \,=\,d(x)/3$ and so we re-obtain eq.~\eqref{eq:fm} - the
total momentum carried by (neutral) particles other than the $u$ and $d$ quarks
is still about $46\%$. The substantial difference between our model and the SM 
is that that percentage of momentum should, in first order of approximation at 
least, be distributed by three gluons only, not eight. The massive gluons are 
very heavy, as we had the opportunity to discuss in section~\ref{sec:glu}, so 
the fraction of momentum they might carry must be extremely small. 
The open question is therefore, can the present results for the 
gluons' momentum distribution functions be reproduced with three massless gluons
instead of eight? 

\subsection{One photon or Z processes}
\label{sec:one}

A quantity that might be used as evidence of FCQ is the ratio of cross sections 
of the processes $e^+\,e^-\,
\rightarrow\,q\,\bar{q}$ and $e^+\,e^-\,\rightarrow\,\mu^+\,\mu^-$, which 
at tree level is
\begin{equation}
R\;=\;\frac{\sigma(e^+\,e^-\,\rightarrow\,q\,\bar{q})}{\sigma(e^+\,e^-\,
\rightarrow\,\mu^+\,\mu^-)}\;\; = \;\; 3\,\sum_i\,Q_i^2 \;\;\; .
\label{eq:rat}
\end{equation}
The $3$ is a colour factor and the sum is over the type of quarks with masses
below the energy of the experiment. Therefore the contribution of an up-type quark to this
ratio is, in the SM, $3\,(2/3)^2 \,=\, 4/3$ and for a down quark, $1/3$.
Now, naively applying this formula to our broken QCD theory with the new quark 
charges we would obtain
\begin{align}
R_u &= \sum_{\mbox colours}\;Q_{u^i}^2\; = \; 1\,+\,1\,+\,0 \;=\; 2 \nonumber \\
R_r &= \sum_{\mbox colours}\;Q_{d^i}^2 \;= \; 0\,+\,0\,+\,1 \;= \; 1 \;\;\; ,
\end{align}
but such calculation would be wrong. With the hadronization hypothesis explained
in the previous section we see that in both the SM and the ICQ theory whenever a
quark of a given colour is produced in a $e^+e^-$ collision what is observed 
after hadronization is a superposition of colourless bound states. Whichever the
final state we conclude that the probability of ``finding" a quark of a 
particular colour is $1/3$. Since individual colours are not observed the total 
amplitude of the process $e^+e^- \,\rightarrow\, q\,\bar{q}$ is the sum of the 
amplitudes for each colour,
\begin{equation}
T(e^+e^- \,\rightarrow\, q\,\bar{q}) \;=\; \sum_{i=1}^3 \; T(e^+e^- \,
\rightarrow\, q^i\,\bar{q}^i) \;\;\; .
\end{equation}
The cross section $\sigma(e^+e^- \,\rightarrow\, q\,\bar{q})$ is therefore 
proportional to the square of the modulus of this amplitude multiplied by a 
factor of $1/3$. Hence we obtain
\begin{align}
R^{SM}_u &= \; \frac{1}{3}\,\left|\frac{2}{3}\,+\,\frac{2}{3}\,+\,\frac{2}{3}
\right|^2 \;= \; \frac{4}{3} \nonumber \\
R^{SM}_d &= \; \frac{1}{3}\,\left|-\frac{1}{3}\,-\frac{1}{3}\,-\frac{1}{3}
\right|^2 \;= \; \frac{1}{3} \;\;\;,
\end{align}
which are the results given by eq.~\eqref{eq:rat}. Applying the same reasoning
to the broken QCD case we obtain
\begin{align}
R_u &= \; \frac{1}{3}\,\left|1\,+\,1\,+\,0\right|^2 \;= \; \frac{4}{3} 
\nonumber \\
R_d &= \; \frac{1}{3}\,\left|0\,+\,0\,-\,1 \right|^2 \;= \; \frac{1}{3} 
\;\;\; ,
\end{align}
and so both theories predict exactly the same value for $R$. This quantity is 
not experimental evidence of fractional charges, but it is irrefutable 
demonstration for the existence of colour. Looking at the new definition of 
electric charge, eq.~\eqref{eq:nQ}, we understand how both theories can lead to 
the same result for $R$ - when we calculate the amplitude of $e^+e^- \,
\rightarrow\, q\,\bar{q}$ by summing over the colours of the final state we are
effectively taking the trace of $\lambda_8$, which is zero, and are thus left 
with the SM result. For $q\,\bar{q}$ production via a single $Z$ boson the 
same argument applies: a factor of $1/3$ multiplies the squared amplitude so 
that the cross section will be 
\begin{align}
\sigma(\ldots \tilde{Z} \rightarrow q\bar{q}) &\sim \; \frac{1}{3}\,\left| 
\sum_i \, {\cal M}(\ldots \tilde{Z} \rightarrow q^i\bar{q}^i)\right|^2 \;=\; 
\frac{1}{3}\,\left( \left|3\,g^{SM}_{Z_{V_q}}\right|^2\,+\,\left|3\,
g^{SM}_{Z_{A_q}} \right|^2 \right) \nonumber \\
 &= \; 3\,\left( \left|g^{SM}_{Z_{V_q}}\right|^2\,+\,\left|g^{SM}_{Z_{A_q}}
\right|^2 \right) \;\;\; ,
\end{align}
which is exactly the SM result. Having chosen the vevs of the theory so that
the $Z$ couplings to both quarks and leptons are identical to the SM tree-level
ones, we find that both theories have identical (tree-level) cross sections
for all processes involving a single $Z$, or indeed a single photon.  

\subsection{Two-photon physics}
\label{sec:icq}

There are two experimental results usually presented against ICQ theories, the  
width of the decay of the pseudoscalar mesons $\eta$ and $\eta^\prime$ into
two photons and the cross section of the process $e^+ e^- \rightarrow e^+ e^- 
\,+\, \mbox{hadrons}$. They have in common the fact they are both processes
involving two photons and we wish to argue that this evidence against ICQ is not
conclusive. For $\eta , \eta^\prime  \rightarrow \gamma \gamma$ a simple PCAC 
analysis yields for the width the expression~\cite{berg}
\begin{equation}
\Gamma_{\gamma\gamma}^X \;=\; \left(\sum_{\mbox{colour}} \,<e_q^2>\right)^2\, 
\frac{\alpha^2}{32\pi^3}\,\frac{m_X^3}{f_X^2} \;\;\;\ ,
\end{equation}
where $\alpha$ is the fine structure constant, $m_X$ and $f_X$ the meson's mass
and decay constant and $e_q$ the charge of its constituent quarks. The colour
sum in the quarks' squared charges yields for the pion, for example (see 
eq.~\eqref{eq:pi0}), a factor of $(3\,(4/9 \,-\, 1/9)/\sqrt{6})^2 \,=\, 1/6$ for
the SM; for our ICQ theory we would have $(1\,+\,1\,-\,1)^2/6\,=\, 1/6$ - the 
same result, that is. For the $\eta_8$ meson - with a flavour wave function 
given by $\eta_8\,=\,(u\bar{u}\,+\,d\bar{d}\,-\,2s\bar{s})/ \sqrt{6}$ - both 
theories give the same result, $1/18$, but for the flavour-singlet state, 
$\eta_1\,=\,(u\bar{u}\,+\,d\bar{d}\,+\,s\bar{s})/ \sqrt{3}$, the SM gives 4/3, 
the ICQ theory 16/3. The ICQ width is therefore 4 times larger than the SM one
{\em if} one assumes nonet symmetry, that is, $f_1 = f_8 = f_\pi$. The physical 
mesons $\eta$ and $\eta^\prime$ being a mixing of $\eta_1$ and $\eta_8$ this
seeming discrepancy was used as evidence against ICQ theories. 

But from very early on it was recognised~\cite{chan} that this argument relied 
too heavily on untested theoretical grounds, that is, the ``nonet symmetry" 
hypothesis; if we had $f_1 \,=\, 2\,f_8$ ICQ theories would be favoured over FCQ
ones. Equality of the mesons' decay constants is tantamount to assuming equality
of the wave functions of $\eta_1$ and $\eta_8$ at the origin, which would make 
sense if the $\eta$ and $\eta^\prime$ were ideally mixed like the $\omega$ and 
$\varphi$. But the $\eta$ and $\eta^\prime$ are far from ideally mixed, the 
difference in the $\eta_1$, $\eta_8$ binding energies being of the order of 
their masses. In fact more elaborate calculations found substantial differences 
between the several meson decay constants. For instance in ref.~\cite{ben1}, 
using a Hidden Local Symmetry model~\cite{ban}, it was found that $f_1 \,=\,1.4
\,f_8$. Nonet symmetry is thus not a trustworthy tool. Chanowitz~\cite{chan} 
deduced equations for a $\xi$ parameter ($\xi = 1$ for FCQ, $\xi = 2$ for ICQ) 
in terms of experimental quantities obtained from the decays $\eta/\eta^\prime 
\rightarrow \gamma \gamma/\pi \pi \gamma$, equations in principle not dependent 
on nonet symmetry. A first evaluation of $\xi$ clearly favoured FCQ. However 
Chanowitz's equations, from the very start, were less reliable if the underlying
theory being tested was ICQ (a consequence of the existence of charged gluons
in ICQ theories that might make for substantial mixing between glueballs and 
pseudoscalar mesons), other than displaying a strong dependence on the 
properties of the $\rho$ meson. Further, more thorough studies~\cite{ben2} 
showed the Chanowitz equations actually favoured ICQ models over a QCD-type one.
Further, the best agreement between the FCQ theory and the experimental data
required the introduction of an {\em outr\'e} mass-dependent non-vector meson
dominance $\gamma \rightarrow \pi^+\pi^-$ contribution. This was all the more 
impressive due to the fact that a separate analysis, independent of the $\rho$
contribution~\cite{afn} led to the same conclusions. This seeming embarrassment
for QCD did not survive more precise measurements of the decay widths of 
$\omega$, $\varphi$ mesons, shown to be in agreement with FCQ over the ICQ 
theory. Finally the analysis of ref.~\cite{ben3} concluded that the failure
of the Chanowitz equations was not necessarily a failure of QCD - rather, that
the equations were being misused and were not conclusive. To add to the immense
complexity of the problem is the uncertainty of the gluonic content of the 
$\eta$, $\eta^\prime$ mesons, with different groups claiming evidence 
for~\cite{ball} or against it~\cite{ben3}. In short a simplistic analysis of the
$\eta/\eta^\prime \rightarrow \gamma \gamma$ decays does provide arguments 
against ICQ theories but more detailed studies show such clear-cut conclusions 
are difficult to establish. We would also add that the theory in discussion in
this paper (three massless gluons four charged and one neutral massive ones) is 
quite different from those tested in refs.~\cite{chan,ben2}. 

\begin{figure}[htb]
\epsfysize=11cm
\centerline{\epsfbox{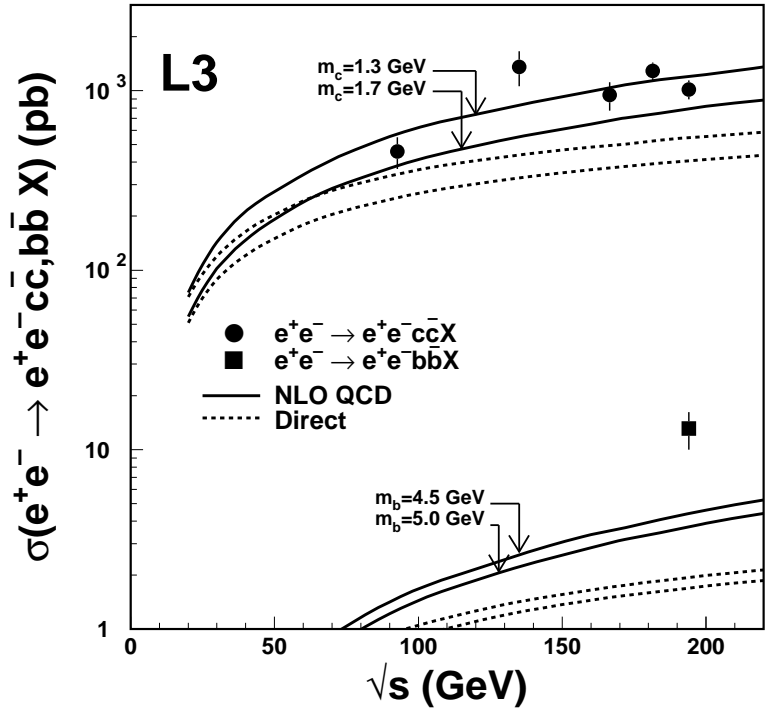}}
\caption{Cross section for production of $c\bar{c}$ and $b\bar{b}$ pairs at
LEP through the two photon channel~\cite{L3q}.}
\label{fig:gg}
\end{figure}

The second experimental evidence rallied against ICQ theories is the cross 
section of the process $e^+ e^- \rightarrow e^+ e^- \,+\, \mbox{hadrons}$ via 
the two-photon channel. As emphasised by Witten~\cite{wit} this process is 
preferential to establish integer/fractional quark charges as its cross section
is proportional to their fourth power - the cancellation of the colour 
contributions observed in section~\ref{sec:one} for the photon and $Z^0$ no 
longer occurs. Defining the ratio
\begin{equation}
R_{\gamma\gamma} \; =\; \frac{\sigma(e^+ e^- \rightarrow e^+ e^- \,+\, 
\mbox{hadrons})}{\sigma(e^+ e^- \rightarrow e^+ e^- \mu^+ \mu^-)} 
\end{equation}
we find that, at tree level, it is given by
\begin{equation}
\left\{ \begin{array}{rl}
\mbox{FCQ}\;\; : & R_{\gamma\gamma}^F \;=\; 3\,\displaystyle{\sum_q}\,e_q^4 \\
\mbox{ICQ}\;\; : & R_{\gamma\gamma}^I \;=\; \displaystyle{\frac{1}{3}}\,
\displaystyle{\sum_q} \,\left( \displaystyle{\sum_{\mbox{i=colours}}}\,
{e_q^i}^2\right)^2 
\end{array} \right. \;\;\; .
\label{eq:rgg}
\end{equation}
So for an up-type quark we have $R_{\gamma\gamma}^I/R_{\gamma\gamma}^F \,
=\, 9/4$ and for a down-type one, $R_{\gamma\gamma}^I/R_{\gamma\gamma}^F \,=\, 
9$. This is, however, a purely perturbative analysis. Again the problem is much
more complex as there is the possibility that the photon-photon interaction
is not perturbative - the so-called ``direct" processes - but rather a 
``resolved" process. At low values of their transverse momentum the photons do 
not interact like point particles but instead like systems with structure,
that is, the photons fluctuate first into an hadronic state and the interaction 
occurs via a gluon or quark from ``inside" the photon~\cite{wit,wal}. For 
reviews see~\cite{berg, kla}. The study of $\gamma\gamma$ interactions at the 
low energy regions depends thus on photon form factors and is quite complicated
requiring Reggeon and Pomeron~\cite{pom} parametrisations to fit, for instance, 
the total cross section of $\gamma\gamma \rightarrow 
\mbox{hadrons}$~\cite{opal1}. These procedures have been successful in 
reproducing - most - experimental results up to LEP2 energies~\cite{wen}. The 
immediate question to ask, though, is this: up to what value of typical 
transverse momentum should the non-perturbative processes be ``needed" to 
describe the experimental data? We remark, for instance, that the $\gamma 
\gamma$ results from PETRA were explained with a perturbative ICQ theory with 
eight massive gluons of mass $\sim 0.3$ GeV~\cite{god}, this for $P_T \sim 5$ 
GeV, though this match may have been accidental. Restating the problem,
when should we compare data with the perturbative predictions alone and thus 
infer their validity? A reasonable expectation would be that the $\gamma
\gamma$ production of heavy quarks would be well described by perturbative QCD, 
particularly in the case of the bottom quark. However, as we may observe in 
fig.~\ref{fig:gg}, although $c\bar{c}$ production is well described by Next to 
Leading Order (NLO) QCD, including both direct (perturbative) and resolved 
processes, $b\bar{b}$ production clearly is {\em not}~\footnote{The high-$P_T$ 
production of $\pi^0$'s in $\gamma \gamma$ reactions suffers from the same
problem, see for example ref.~\cite{L3}.}, the theoretical prediction being a 
factor of 3 or so below the experimental result. In both cases, though, we see 
that the direct processes - the type A diagrams of ref.~\cite{dree} - are not 
enough to describe the data, {\em for an FCQ theory}. The $b\bar{b}$ result has 
been used to argue for the existence of new physics, with supersymmetry brought 
into play to account for it~\cite{ber} via a light colour octet gluino. In 
reference~\cite{icq} another interpretation was proposed, that these results 
are caused by the quarks having integer electric charges. Let us admit that the 
cross sections we are observing at these high energies are purely perturbative 
ones, an expectation that does not seem absurd given the high masses of the 
charm and bottom quarks. To obtain the ICQ predictions, the FCQ direct cross 
section for $c\bar{c}$ production should be multiplied by a factor of $9/4$ and 
the $b\bar{b}$ one by a factor of $9$ (see eq.~\eqref{eq:rgg}). That this should
constitute a very good approximation to the equivalent ICQ cross sections may be
seen from the explicit expressions for $\sigma^{FCQ}(\gamma\gamma \rightarrow Q
\bar{Q})$ from ref.~\cite{dree}. Both the pure QED and the QCD corrections are 
proportional to the fourth power of the quarks' charges so the factors we 
described make the transition from a FCQ to an ICQ scenario. The strong 
interaction corrections are however clearly different in the SM and our ICQ 
theory. For starters we do not expect the massive gluons to have substantial 
contributions to these processes, given their large masses. Therefore an 
approximation to the QCD corrections in our theory should involve only three 
massless gluons and the factor of $4/3$ in expression (2) of ref.~\cite{dree} 
should be replaced by the appropriate $SU(2)$ Casimir, $3/4$. It is thus trivial
to see that our simple approximation to $\sigma^{ICQ} (\gamma\gamma \rightarrow 
Q\bar{Q})$ overestimates the QCD corrections. However, also from 
ref.~\cite{dree}, we see that for high values of the center-of-mass energies the
$\alpha_s$ corrections are expected to be substantially smaller than the QED 
contributions.  Having thus argued that $9/4 \,\times\,\sigma^{FCQ}(\gamma\gamma
\rightarrow c\bar{c})$ should be a good approximation to $\sigma^{ICQ}(\gamma
\gamma \rightarrow c \bar{c})$ and $9\, \times\,\sigma^{FCQ} (\gamma\gamma 
\rightarrow b\bar{b})$ to $\sigma^{ICQ}(\gamma\gamma \rightarrow b\bar{b})$ we 
gather the experimental results from LEP2 at 194 GeV from the L3 
collaboration~\cite{L3q} (entirely compatible with the results from the OPAL 
collaboration~\cite{opal2} and with more recent data from L3~\cite{L3n}) and 
their direct FCQ predictions. We find the results shown in table~\ref{tab:cros}.
\begin{table}[t]
\begin{center}
\begin{tabular}{|c|c|c|c|} \hline & & & \\
 & FCQ Direct & ICQ Direct & Exp. Data  \\ & & & \\ \hline & & & \\
$\sigma(\gamma\gamma \rightarrow c\bar{c})$ & 400 - 525 & 900 - 1181.3 & $1016 
\pm 123$ \\ & & & \\ \hline & & & \\ 
$\sigma(\gamma\gamma \rightarrow b\bar{b})$ & 1.54 - 1.78 & 13.9 - 16.0 & 
$13.1 \pm 3.1$ \\ & & & \\ \hline
\end{tabular}
\caption{Comparison of experimental data and FCQ and ICQ direct (perturbative)
NLO QCD for $c\bar{c}$ and $b\bar{b}$ production cross sections. All results
are in picobarns. The range in the theoretical predictions comes from varying
the value of the quarks' mass ($1.3 \leq m_c \leq 1.7$ GeV for $\sigma(\gamma
\gamma \rightarrow c\bar{c})$, $4.5 \leq m_b \leq 5.0$ GeV for $\sigma(\gamma
\gamma \rightarrow b\bar{b})$.), the lower cross sections corresponding to 
higher masses.}
\end{center}
\label{tab:cros}
\end{table}
We argue that the agreement between the ICQ approximate predictions and the LEP2
data is compelling enough to seriously consider the possibility that the
quarks do have integer charges~\footnote{This simple approximation to the ICQ
cross section does not explain the $\pi^0$ production cross section excess
found at LEP~\cite{L3}, but given the low mass of the up and down quarks one 
may argue the perturbative results cannot ``yet" be trusted in that case.}. The 
agreement between the ICQ prediction and the $c\bar{c}$ data is not an accident
for this particular collision energy, it occurs at energies ranging from $91$ to
$194$ GeV, as can be seen in ref.~\cite{icq}. One cannot forget, however, that 
the data from ref.~\cite{L3q} were obtained assuming the experimental 
backgrounds were those dictated by the SM. A complete comparison of data to our 
ICQ predictions would have to take into account the different QCD gauge group. 
Also worthy of note in the $ b\bar{b}$ case is that, as we already mentioned, 
the bottom quarks mix with the charginos in this CCB scenario. This simple 
analysis doesn't take that into account - that mixing, though, depends on the 
supersymmetry-breaking parameters and thus one can expect sufficient freedom to 
reproduce the data. That complete analysis could be quite interesting in that it
would presumably constrain the MSSM parameter space.  

\section{Discussion and conclusions}
\label{sec:ana}

According to the Particle Data Group~\cite{pdg} the current limit on the mass of
the gluon is ``a few MeV". That, however, is a {\em theoretical} bound, derived
from arguments that assume all gluons are degenerate in mass~\cite{ynd}. That 
is most definitely not the case of the theory presented in this paper. Also, as 
others have discussed~\cite{fiel}, the arguments of~\cite{ynd} neglect to take
into account the quantum field theoretical aspects of the gluon-quark  
interactions, which casts doubt over their final conclusions. Many theories have
been developed where gluons acquire mass by means of gauge symmetry 
breaking~\cite{pati,glu} or the introduction of a four-vertex ghost 
field~\cite{fad}. Cornwall proposed a mechanism for dynamic generation of 
gluon masses within a theory with unbroken colour gauge symmetry~\cite{corn}.
Recently Field~\cite{fiel} considered gluons of mass $\sim 1$ GeV to explain the
radiative decays of the $J/\Psi$ and $\Upsilon$ mesons. His is a convincing 
argument for the existence of massive gluons. Of course, the gluons discussed in
this paper have masses much larger than Field's, and our theory also includes 
three massless ones. Therefore it remains to be seen if the analysis 
of~\cite{fiel} could be reproduced by our ICQ model.  

We obtain ICQ from charge and colour breaking in a supersymmetric theory, though
to be precise we are not {\em really} breaking electric charge symmetry - as we 
see from formula~\eqref{eq:nQ} the vevs $v$, $q$ and $t$ are neutral for the new
definition of electric charge. For the old one, though, they would appear as 
charged. Now, we were able to determine the values of these vevs by looking at 
the electroweak couplings and requiring them to equal the corresponding SM 
values. For this scenario to be possible the SUSY-breaking parameters must be 
such that a CCB minimum is produced in the potential. It has been 
established~\cite{cas} that CCB associated with the top Yukawa coupling is 
difficult to obtain (because the $SU(3)$ D-term contributions to the potential 
are large and positive). One-loop studies of CCB potentials~\cite{tau,tau2} 
revealed commonly accepted CCB bounds could be overestimated. Considering these 
restrictions it is then possible that only a very small portion of SUSY 
parameter space could produce the pattern of CCB we are interested in. This is 
an interesting feature of this model as it could increase immensely the 
predictive power of the MSSM. Notice that CCB bounds have been considered to 
avoid the possibility of CCB occurring, whereas in such a study we would want 
to enforce it. We hope to address this subject in the future. 

\begin{figure}[htb]
\epsfysize=9cm
\centerline{\epsfbox{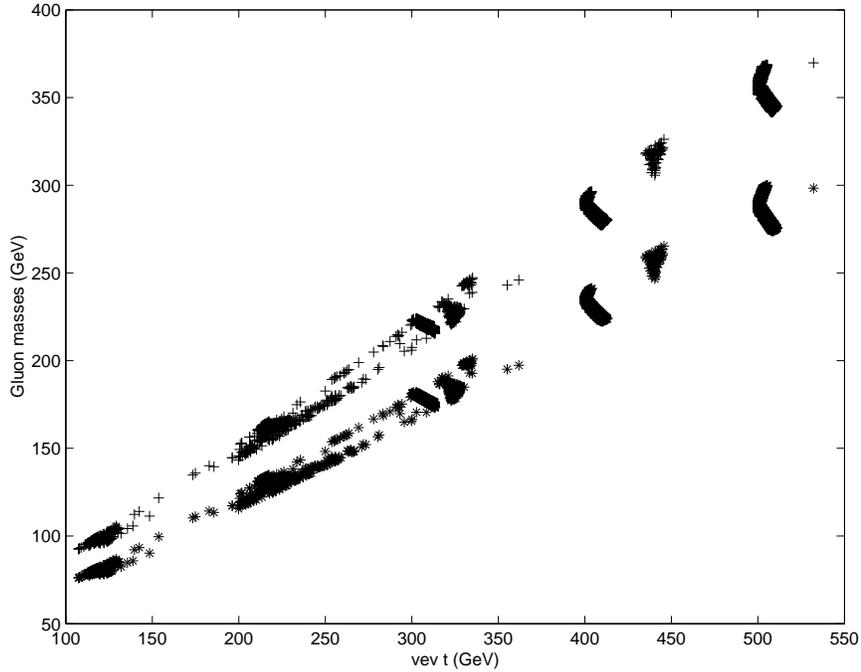}}
\caption{Masses of the gluons $\tilde{G}$ (crosses) and $G$ (stars) {\em vs.} 
the vev $t$, for different initial guesses for $t$, 100, 200, 300, 400 and 500 
GeV.}
\label{fig:mgl}
\end{figure}

We have built a model that closely mimics the electroweak sector of the SM but
its strong sector is very much an open question. It remains to be seen whether
our spectrum of gluons is able to reproduce the plethora of QCD experimental 
results in existence. For instance, proton-proton collisions. We would expect 
that, given the large mass of five of our gluons, only the massless ones would 
participate in $p-p$ processes. The partonic cross sections for those reactions
would therefore not be the usual ones. But then we have to remember 
(section~\ref{sec:had}) that the gluons' distribution functions are also 
different - we now have, in first approximation at least, three gluons carrying
about $46\%$ of the proton's momentum, not eight like in the SM. The observed
cross section being the convolution of the partonic cross sections and the 
parton distribution functions, the possibility exists that the end result, 
amazingly, be the same for both theories. There are however two experimental
results that challenge the notion of an unbroken $SU(2)$ colour gauge group. The
first is the measurement of the gluon and quark fragmentation functions at 
DELPHI and determination of the ratio $C_A/C_F$~\cite{caf}. For a gauge group
$SU(N)$ this ratio is given by $C_A/C_F \,=\, 2\,N^2/(N^2-1)$. The result 
from~\cite{caf} is $2.26\pm 0.16$, in agreement with the predicted value for
$SU(3)$, $9/4$ - but not too far from the $SU(2)$ result, $8/3$. We have to 
remark that the analysis of ref.~\cite{caf} is strongly dependent on the jet 
algorithm used and the choice of energy scale. Also, the event selection in
that work was based on the expected behaviour of three-jet systems in an 
unbroken $SU(3)$ theory. Would the results be different, we ask, if the 
underlying theory was assumed to be an unbroken $SU(2)$ with massive gluons? 
A more serious objection is the one posed by the measurements of the strong
coupling $\alpha_s$ at low energies~\cite{pdg} - the evolution of $\alpha_s$
with the energy scale is very well described by a SM $SU(3)$ $\beta$-function
with the appropriate number of quark flavours. However, it is possible that 
admitting that the $\beta$-function for our ICQ model below, say, the $M_Z$ 
scale, is equivalent to that of an unbroken colour $SU(2)$ group is an 
oversimplification. In fact, as we can appreciate from fig.~\ref{fig:mgl}, for 
masses of $\tilde{G}$ as high as $\sim 130$ GeV, we have charged gluon masses 
lower than $M_Z$. In such a situation we would expect those gluons to have a 
contribution to the running of $\alpha_s$ near the weak scale, at least. We must
also remember that the evolution of $\alpha_s$ should go accordingly to a 
supersymmetric $\beta$-function. These reduce to the SM ones at energy scales 
below the mass of the lightest supersymmetric partner (when all sparticles have 
been ``integrated out" of the theory and the spectrum remaining is that of the 
SM). If we have a spectrum with sparticle masses lower than $M_Z$ (the 
experimental bounds on sneutrino or neutralino masses, for instance, are still 
well below $M_Z$) we would expect the $\alpha_s$ running of such a theory to be 
more elaborate that that dictated by a simple SM $SU(2)$ $\beta$-function. The
scale at which we should start assuming that the running of $\alpha_s$ is given 
by a ``pure" $SU(2)$ $\beta$-function is thus dependent on the particular SUSY 
spectrum we choose. Also, particle threshold corrections can have significant 
impact on the low-energy value of the coupling constants~\cite{dam}. This means 
further study is necessary to determine if the theory is capable or not of 
reproducing the low energy running of the strong coupling constant. 

In conclusion, we built a theory with broken colour gauge symmetry. The vevs in
this theory, according to SM expectations, have electric charge and thus the 
photon was expected to become massive. We saw that this isn't so as the 
symmetry breaking leads to a new definition for the electric charge, with an 
extra contribution from $SU(3)$. This new definition does not affect the lepton
charges but changes the quarks' - they become dependent on the colour of 
the quark in question and are integer. As a consequence of gauge symmetry 
breaking five gluons gain mass, four of those are electrically charged. The 
neutral massive gluon is mixed with the photon and $Z$ fields, this mixing 
causing the appearance in the theory of new vertices, such as $\tilde{G}\,e^+\,
e^-$. This ICQ model closely mimics the electroweak sector of the SM - it is 
possible to choose the vevs such that the couplings of the new photon and $Z^0$ 
are equal to the tree-level values of the same particles in the SM. The mass and
interactions of the $W$ gauge boson are also unchanged, except for the new
gauge boson self-interaction vertices mentioned earlier. The mass of the gluons
was found to be strongly dependent on the choice of a particular vev - the $t$
vev, associated with the scalar field $t_R$ - and very high. In the electroweak
sector the two theories have different experimental predictions only in 
two-photon processes. We have argued that recent LEP2 results are quite 
naturally explained admitting quarks have integer charges, and we must recall
that the SM prediction for the cross section for $\gamma \gamma \rightarrow 
b\bar{b}$ is a factor of 3 below the experimental result. There are however 
experimental results that would seem to contradict the idea of an unbroken 
$SU(2)$ colour gauge group, the measurements of $C_A/C_F$ and the low energy
running of $\alpha_s$. In both cases it was argued that there is a strong model
dependence in the analysis of the data or the ICQ predictions. As such, the 
theory may still be able to accommodate these results. On theoretical grounds 
the weakest point in this theory seems to be the hadronization hypothesis, that 
hadronic bound states are singlets of $SU(3)$ even though that symmetry is now 
broken. As discussed, this seems no more outrageous than the SM confinement 
scheme, given our degree of ignorance of the non-perturbative physics that 
governs hadronization. Notice that we tested this model at tree-level only. The 
theory is obviously renormalisable so higher order corrections will be small. 
Given the degree of accuracy achieved at LEP loop corrections to the ICQ model
may prove to be a good testing ground of the theory. Another experimental 
prediction is the interesting charge distribution in hadrons presented in 
section~\ref{sec:had}, that we hope to look into in the near future. The
existence of the massive gluon $\tilde{G}$, with direct coupling to leptons, 
would of course be ideally tested in an electron-positron linear collider of
high energy, like the TESLA or NLC projects. In short, and to answer this 
article's title question, it seems possible to build a theory with broken charge
and colour symmetries with sensible behaviour in the electroweak sector, at 
least. The strong sector will be the hardest test of this theory. It will be 
necessary to show if the different gluon spectrum can reproduce the wealth of
experimental QCD data existent. The mixing of quarks, neutralinos, charginos and
gluinos existing in this model should also be carefully studied as both a 
possible window into the supersymmetric sector of the theory or a means of 
excluding it, if the changes in the bottom or top mass/interactions are too
severe. The interest in considering such a theory as an alternative to the SM?
If the $\gamma\gamma \rightarrow b\bar{b}$ result from LEP2 cannot truly be
explained by the SM new physics will be necessary. ICQ theories were shown to 
fit admirably well that data - and the ICQ theory presented in this work is
the simplest one there can be, as it requires a gauge group no larger than the
SM's. It also provides new insight into the supersymmetric sector and, from
the CCB requirement, we hope it will have a reduced allowed parameter space and
thus considerable predictive power. 

\vspace{0.25cm}
{\bf Acknowledgments:} this work was made possible through extensive 
discussions with prof. Augusto Barroso for which, many thanks. For their 
suggestions, criticism and help I thank Pedro Abreu, Ant\'onio Amorim, Sofia 
Andringa, Carlo Becchi, Jorge Dias de Deus, Tim Jones, M\'ario Pimenta and 
Giovanni Ridolfi. This work was supported by a fellowship from Funda\c{c}\~ao 
para a Ci\^encia e Tecnologia, SFRH/BPD/5575/2001.

\end{document}